\documentclass[twocolumn,showpacs,aps,floatfix]{revtex4}

\usepackage{amsmath}
\usepackage{graphicx}
\usepackage{dcolumn}
\usepackage{bm}

\begin{document}
\preprint{ND Atomic Experiment 2004-3}
\title{1s2s2p$^{2}$3d $^{6}$L - 1s2p$^{3}$3d $^{6}$D, L=F, D, P Transitions in
O IV, F V and Ne VI}
\author{Bin Lin$^{1}$}
 \email{blin@nd.edu}
 \homepage{http://www.nd.edu/~blin/}
\author{H. Gordon Berry$^{1}$}
 \email{Berry.20@nd.edu}
 \homepage{http://www.science.nd.edu/physics/Faculty/berry.html}
 \author{Tomohiro Shibata$^{1}$}
 \author{A. Eugene
Livingston$^{1}$}
 \author{Henri-Pierre Garnir$^{2}$}
 \author{Thierry Bastin$^{2}$}
 \author{J.
D\'{e}sesquelles$^{3}$}
\affiliation{1 Department of Physics, University of Notre Dame, Notre Dame, IN 46556\\
2 IPNAS, University of Liege, B4000 Liege, Belgium \\
3 Lab Spectrometrie Ion \& Mol, University of Lyon, F-69622 Villeurbanne, France}
\date{\today}
\begin{abstract}
We present observations of VUV transitions between doubly excited
sextet states in O IV, F V and Ne VI. Spectra were produced by
collisions of an O$^{+}$ beam with a solid carbon target. We also
studied spectra obtained previously of F V and Ne VI. Some
observed lines were assigned to the 1s2s2p$^{2}$3d $^{6}$L -
1s2p$^{3}$3d $^{6}$D, L=F, D, P electric-dipole transitions, and
compared with results of MCHF (with QED and higher-order
corrections) and MCDF calculations. 42 new lines have been
identified. Highly excited sextet states in five-electron ions
provide a new form of energy storage and are possible candidates
for VUV and x-ray lasers.
\end{abstract}

\maketitle
\section{Introduction}

Highly excited sextet states in five-electron ions provide a form
of energy storage. The research for stimulated VUV- and x-ray
emission from highly excited sextet states in five-electron ions
has attracted attention in recent years. This new form of energy
storage and potential VUV-ray lasers could have many applications
in basic science, technology, medicine, and defense. In a proposed
VUV and x-ray laser system, one seeks a probability to trigger a
release of K- and L-hole energies of sextet states in boron-like
ions. Although K- and L-hole energies are not as high as energies
released in nuclear fusion, capacity of highly excited sextet
states in five-electron ions to store energy is significant
(several hundred electron volts per atom shown in Fig. 1). Such a
system involves long-lived ''storage'' metastable states, and
there nearby are short-lived higher excited sextet states, from
which transitions are emitted with photon radiation, and quintet
continuum. An ideal system where VUV and x-ray lasers may be
implemented would be among heavy and highly excited ions that have
metastbale sextet states with long lifetimes. However, structure
and transition properties of these sextet states are currently
very poorly known.
\begin{figure}[tbp]
\centerline{\includegraphics*[scale=0.85]{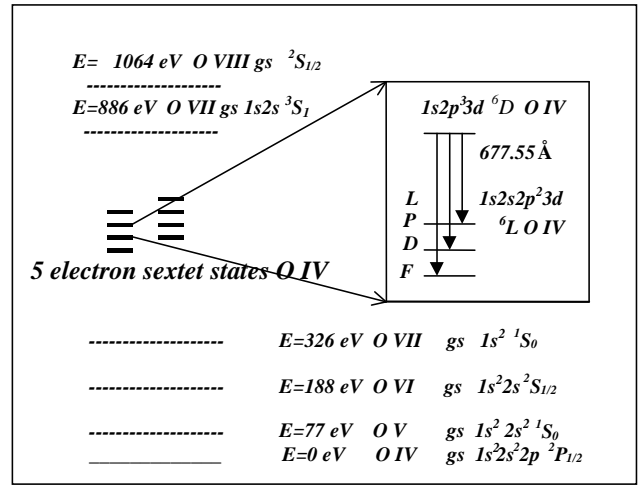}} \caption{Term
diagram of doubly excited sextet states of O IV.
The mean wavelength for the 1s2s2p$^{2}$3d $^{6}$L-1s2p$^{3}$3d $^{6}$D%
$^{o}$ transitions in O IV is shown.} \label{fig1}
\end{figure}

In 1992 beam-foil spectroscopy~\cite{bl,lap} was used to provide
initial data on low-lying sextet states in doubly excited
boron-like nitrogen, oxygen and fluorine. Recent work of Lapierre
and Knystautas~\cite{lap} on possible sextet transitions in Ne VI
highlights the significance in this sequence. They measured
several excitation energies and lifetimes. Fine structures of
individual 1s2s2p$^{2}$3s $^{6}$P$_{J}$ states were resolved and
measured in O IV, F V and Ne VI by Lin and Berry et al~\cite{lin}.
There are no further results reported for transitions from highly
excited sextet states.

In his work, fast beam-foil spectra of oxygen were recorded at
Liege using grading incidence spectrometers~\cite{berry1, kla}.
Spectra of fluorine and neon were previously recorded at the
University of Lyon and the Argonne National Lab,
The 1s2s2p$^{2}$3d $^{6}$L - 1s2p$^{3}$3d $%
^{6}$D, L=F, D, P electric-dipole transitions in O IV, F V and Ne
VI have been searched in these spectra, and compared with results
of MCHF (with QED and higher-order corrections) and MCDF
calculations.

\section{THEORY}
Energies, lifetimes and relevant E1 transition rates of the
doubly excited sextet states 1s2s2p$^{2}$3d $^{6}$L, L=F, D, P and 1s2p$^{3}$%
3d $^{6}$D in boron-like O IV, F V and Ne VI were calculated with
Multi-configuration Hartree-Fock (MCHF) method~\cite{mie,fibk}
(with QED and higher-order relativistic
corrections~\cite{lin,kt1}), and Multi-configuration Dirac-Fock
(MCDF) GRASP code~\cite{MCDF, MCDF1, MCDF2}.

For a sextet state in a five-electron system (\ss , LS=5/2JM$_{J}$)=(n$_{1}$l$%
_{1}^{w1}$n$_{2}$l$_{2}^{w2}$ n$_{3}$l$_{3}^{w3}$ n$_{4}$l$_{4}^{w4}$ n$_{5}$%
l$_{5}^{w5}$ $^{6}$L$_{J}$, M$_{J}$ ), where wi=0,1, ..., or min
(2l$_{i}$+1), i=1,2,... 5, the wavefunction is
\begin{equation}
\Psi (\beta , LS=5/2J)=\sum\limits_{i=1}^{N}\sum\limits_{M_{J}=-J}^{J}c
_{i} \phi (\beta _{i},LS=5/2JM_{J}),
\end{equation}
where c$_{i}$ is a configuration interaction coefficient, N is
a total number of configurations with the same LSJM$_{J}$ and parity, and $\phi $(\ss $_{i}$%
,LS=5/2JM$_{J}$) is a configuration state function (CSF).

In single-configuration Hartree-Fock (SCHF) calculations only the
configurations corresponding to the desired levels, 1s2s2p$^{2}$3d
or 1s2p$^{3}$3d, were considered. After updating MCHF codes we
performed relativistic calculations with an initial expansion of
up to 4000 CSFs and a full Pauli-Breit Hamiltonian matrix. For a
five-electron system a CI expansion generated by an active set
leads to a large number of expansions. To reduce the number of
configurations, we
chose configurations n$_{1}$l$_{1}$n$_{2}$l$_{2}$ n$%
_{3}$l$_{3}$ n$_{4}$l$_{4}$ n$_{5}$l$_{5}$, where n$_{i}$=1, 2, 3,
4 and 5, l$_{i}$=0, ...min (4, n$_{i}$-1). We did not include g
electrons for n=5 shell. For MCHF calculations of the lower states
1s2s2p$^{2}$3d $^{6}$L, L=F, D, P we chose 1s, 2s, 2p, 3s, 3p, 3d,
4s, 4p, 4d and 5s electrons to compose configurations. For the
1s2p$^{3}$3d $^{6}$D state we chose 1s through 4d electrons. Fine
structure splitting is strongly involved in the experiments and
identifications. After determining radial wavefunctions we
included relativistic operators of mass correction, one- and
two-body Darwin terms and spin-spin contact term in both SCHF and
MCHF calculations; these were not included by Miecznik et
al~\cite{mie}.

We used the screened hydrogenic formula from~\cite{lin,qed1,qed2,
drake} to estimate quantum electrodynamic effects (QED) and
higher-order relativistic contributions for sextet states in
five-electron oxygen, fluorine and neon.

In MCDF~\cite{MCDF, MCDF1, MCDF2} calculations, firstly we used
single-configuration Dirac-Fock approach (SCDF). A basis of
jj-coupled states to all possible total angular momenta J from two
non-relativistic configurations, 1s2s2p$^{2}$3d and
1s2p$^{3}$3d, was considered. For convergence we included the ground state 1s$^{2}$2s$%
^{2}$2p of the five-electron systems. After calculating all
possible levels for all J, eigenvectors were regrouped in a basis
of LS terms. To obtain better evaluations of correlation energies
of the doubly excited sextet terms 1s2s2p$^{2}$3d $^{6}$L, L=F, D,
P and 1s2p$^{3}$3d $^{6}$D in O IV, F V and Ne VI, improved
calculations included 1s$^{2}$2s$^{2}$2p,
1s$^{2}$2s2p$^{2}$, 1s2s$^{2}$2p$^{2}$, 1s2s2p$^{3}$, 1s2s2p$^{2}$3s, 1s2s2p$%
^{2}$3p, 1s2s2p$^{2}$3d, 1s2p$^{3}$3s, 1s2p$^{3}$3p, 1s2p$^{3}$3d, 1s2p$^{3}$%
4s, 1s2p$^{3}$4p and 1s2p$^{3}$4d mixing non-relativistic configurations.

In GRASP code~\cite{MCDF, MCDF1, MCDF2} QED effects, self-energy
and vacuum polarization correction, were taken into account by
using effective nuclear charge Z$_{eff}$ in the formulas of QED
effects, which comes from an analogous hydrogenic orbital with the
same expectation value of r as the MCDF-orbital in
question~\cite{MCDF, MCDF1, MCDF2}.

\section{EXPERIMENT}
The experiments were performed with a standard fast-ion beam-foil
excitation system at a Van de Graaff accelerator beam line at the
University of Liege~\cite{kla,berry2,neq,gar,rob}. To produce
spectra of oxygen in the wavelength region near 660-710 \AA\ a
beam current of about 1.3
$\mu$%
A of $^{32}$O$_{2}^{+}$ and $^{16}$O$^{+}$ ions accelerated to
energies of 1.5 and 1.7 MeV were yielded at the experimental
setup. Such energies were expected to be an optimum for the
comparison and production of O$^{3+}$ ions by ion-foil
interaction~\cite{gir}.

The beam current goes through a carbon exciter foil. The foils
were made from a glow discharge, had surface densities about 10-20
$\mu$g/cm$^{2}$ and lasted for 1-2 hours under the above
radiation.

VUV radiation emitted by excited oxygen ions was dispersed by a
1m- Seya-Namioka grating-incidence spectrometer at about 90
degrees to the ion beam direction. A low-noise channeltron (below
1 count/min) was served as a detector. Spectra were recorded at
energies of 1.5 and 1.7 MeV with 100/100 $\mu$m slits (the line
width (FWHM) was 1.1 \AA ) and 40/40 $\mu$m slits (the line width
(FWHM) was 0.7 \AA ) in the wavelength range of 660-710 \AA .

We have reinvestigated unpublished beam-foil spectra of
$^{16}$O$^{3+}$, $^{19}$F$^{4+}$ and $^{20}$Ne$^{5+}$ ions
recorded previously by accelerating $^{16}$O$^{+}$,
$^{20}$(FH)$^{+}$ and $^{20}$Ne$^{+}$ ions to beam energies of 2.5
MeV, 2.5 MeV and 4.0 MeV at the University of Lyon and the Argonne
National Lab. The line width (FWHM) was 0.3 \AA , 0.8 \AA\ and 0.3
\AA\ in the wavelength range of 660-710 \AA , 567-612 \AA\ and
490-555 \AA\ in the spectra, respectively.

\section{RESULTS}
\begin{figure}[tbp]
\centerline{\includegraphics*[scale=0.82]{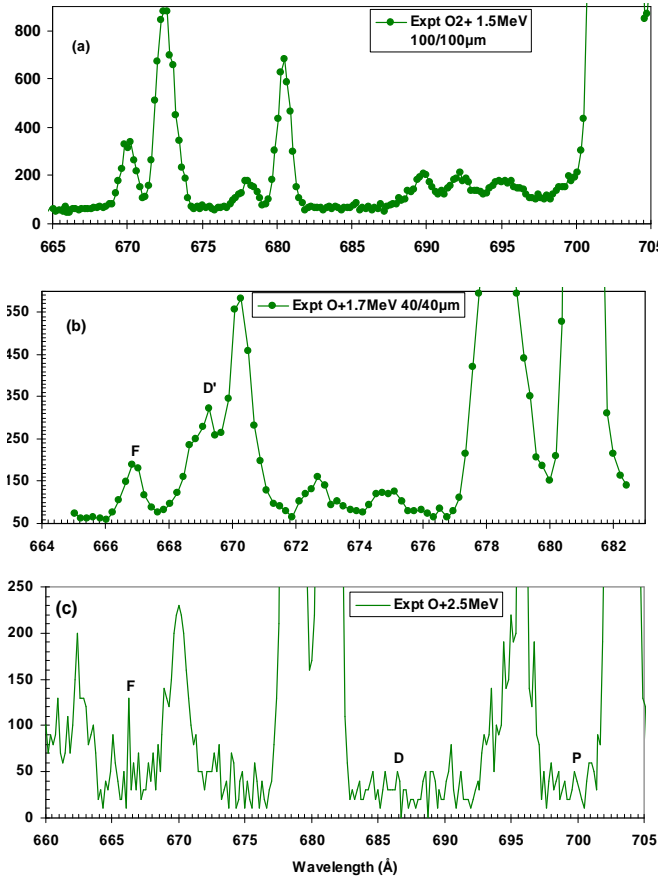}}
\caption{Beam-foil spectra of oxygen. Beam energies and
spectrometer slit widths are indicated. Units of intensities are
arbitrary. F, D and P:
1s2s2p$^{2}$3d $^{6}$L$_{J}$-1s2p$^{3}$3d $^{6}$D$_{J^{\prime }}^{o}$%
, L=F, D and P transitions in O IV. D': 1s2s2p$^{2}$3p
$^{6}$D$_{J}^{o}$-1s2p$^{3}$3p $^{6}$P$_{J^{\prime }}$ transition
in O IV~\cite{lin1}.} \label{fig2}
\end{figure}
Fig. 2(a) -2(c) display three typical spectra of oxygen at beam
energies of 1.5, 1.7 and 2.5 MeV in the wavelength range of
660-710 \AA . In the wavelength region of 660-710 \AA\ transitions
between the sextet states 1s2s2p$^{2}$3d $^{6}$L - 1s2p$^{3}$3d
$^{6}$D, L=F, D, P in O IV were expected. At an O$_{2}$$^{+}$ beam
energy of 1.5 MeV, O$_{2}$$^{+}$ ions are mainly excited to terms
of O$^{2+}$ and O$^{3+}$. There are no lines emitted from sextet
states in O IV in Fig. 2(a). At an O$^{+}$ beam energy of 1.7 MeV,
O$^{+}$ ions are mainly excited to terms of O$^{3+}$ and O$^{4+}$.
New and unidentified emissions appear in the spectrum in Fig.
2(b). Fig. 2(c) shows a spectrum with better resolution to see
details of lines.

For the 1s2s2p$^{2}$3d $^{6}$L - 1s2p$^{3}$3d $^{6}$D, L=F, D, P
transitions we
expected to resolve fine structures of the lower states 1s2s2p$^{2}$3d $%
^{6}$L in the experiments, whereas fine structures of the upper
states 1s2p$^{3}$3d $^{6}$D are close and less than resolution of
the experimental spectra. O V 3p-4d, O IV 2s$^{2}$3p-2s$^{2}$5s, O
V 2s3d-2s4f and O III 2s$^{2}$2p$^{2}$-2s2p$^{3}$ transitions are
at 659.589 \AA , 670.601 \AA , 681.332 \AA\ and 703.854 \AA ,
respectively, close to the neighborhood of the doubly excited
sextet transitions. The four wavelengths have been semiempirically
fitted with high accuracy $\pm $0.004 \AA\ by ~\cite{o1,o2,o3} and
provide a good calibration for the measurements. Standard error
for wavelength calibration is $\pm $0.01 \AA\ in the wavelength
region of 660-710 \AA . Nonlinear least square fits of Gaussian
profiles gave values for wavelengths, intensities and full widths
at half maximum (FWHM) of lines. Uncertainties of wavelengths are
related to intensities of lines. Through the use of optical
refocusing we achieved spectroscopic line width of 0.3 \AA .
Precision of the profile-fitting program was checked through
several known transition wavelengths.

\begin{figure}[tbp]
\centerline{\includegraphics*[scale=1.3]{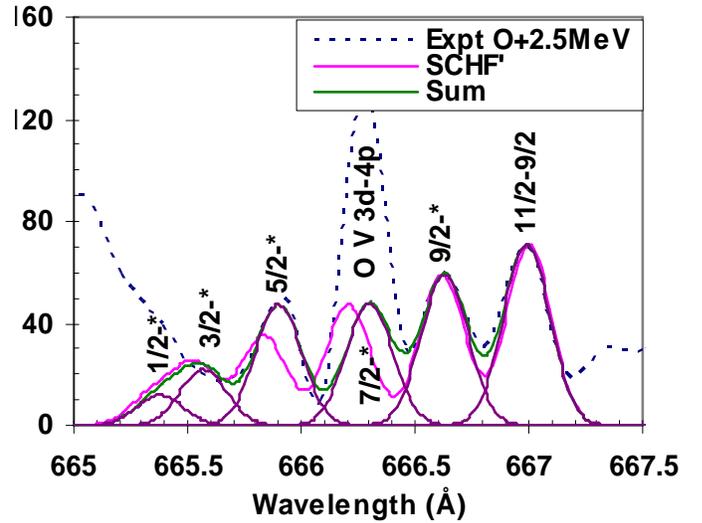}}
\caption{Relative intensity of the 1s2s2p$^{2}$3d $^{6}$F$_{J}$ - 1s2p$%
^{3}$3d $^{6}$D$^{o}_{J^{\prime }}$ transition in O IV in the
experimental spectrum of oxygen at a beam energy of 2.5 MeV. Unit
of intensity is arbitrary. * represents all possible J's of the
upper state allowed by E1 transition rules.} \label{fig3}
\end{figure}

Most of new identifications have been obtained by searching in the
spectra for sets of unidentified lines and by comparing energies
and relative intensities of the 1s2s2p$%
^{2}$3d $^{6}$L-1s2p$^{3}$3d $^{6}$D$^{o}$, L=F, D and P
transitions with results of calculations by MCHF and MCDF
approaches. A promising candidate for the 1s2s2p$^{2}$3d $^{6}$F$%
_{11/2}$-1s2p$^{3}$3d $^{6}$D$_{9/2}^{o}$ transition appears at
the wavelength of 666.99$\pm $0.06 \AA\ in the spectra in Fig.
2(b) and 2(c) recorded at $^{20}$O$^{+}$ ion beam energies of 1.7
and 2.5 MeV, which does not appear in the spectrum in Fig. 2(a)
recorded at
$^{20}$O$_{2}^{+}$ ion beam energy of 1.2 MeV. Shown in Fig. 3 are details of the 1s2s2p$%
^{2}$3d $^{6}$F$^{o}$-1s2p$^{3}$3d $^{6}$D transition in O IV
recorded at an O$^{+}$ ion energy of 2.5 MeV. The curve SCHF' is
convoluted theoretical profile of fine structure components with a
Gaussian function. The experimental width of 0.3 \AA \ for the
oxygen spectrum was utilized. The transition rates to fine
structure j=11/2 to 1/2 of the lower state were results of
single-configuration Hartree-Fock (SCHF) calculations by this
work. The wavelengths of fine structure components were calculated
SCHF results plus a fitted shift for all five components. Measured
wavelength of a component is the weighted center of the fitted
profile of experimental data. Experimental transition rate is
proportional to area of a peak (fitted intensity$\times $FWHM of
experimental data). The curve "Sum" is summation of fitted fine
structure components of experimental data. Ratio of measured
transition rates for J=11/2-9/2, 9/2-*, J=7/2-*, J=5/2-*, J=3/2-*
and J=1/2-* components at an ion energy of 2.5 MeV in Fig. 3 is
about 67.4$\times $0.3:56.3$\times $0.3:45.5$\times
$0.3:46.2$\times $0.28:20.9$\times $0.3:11.3$\times $0.3 =
5.96:4.98:4.02:3.81:1.85:1.00.
* represents all possible j's of the upper state 1s2p$^{3}$3p
$^{6}$P$_{j}$ allowed by electric-dipole transition rules. The
ratio is slightly different from theoretical ratio of E1 GF values
(in length gauge) of SCHF calculations of
1.188:0.990:0.793:0.595:0.397:0.198 =
6.00:5.00:4.00:3.00:2.00:1.00. Based on
above analysis we assign the set of lines as the 1s2s2p$%
^{2}$3d $^{6}$F-1s2p$^{3}$3d $^{6}$D$^{o}$ transition in O IV, and
determine their wavelengths with good accuracy of $\pm $0.06 \AA .
\begin{center}
\begin{table*}
\caption{\label{tab:table1}Energies E (in cm-1) and wavelengths
$\lambda $ (in \AA ) for the 1s2s2p$^{2}$3d $^{6}$L$_{J}$ -
1s2p$^{3}$3d $^{6}$D$_{J^{\prime }}$, L= F, D, P transitions in O
IV by this work. We list differences dE between theoretical and
experimental transition energies for the transitions.}
\begin{ruledtabular}
\begin{tabular}{llrlrrlrrlrrlrr}
J-J' & $\lambda $exp & Eexp & $\lambda $mchf & Emchf & dEmchf &
$\lambda $schf & Eschf & dEschf & $\lambda $mcdf
& Emcdf & dEmcdf & $\lambda $scdf & Escdf & dEscdf \\
  \hline
 & $\pm $0.06 &  &  &  &  &  &  &  &  &  &  &  &  &  \\
\multicolumn{4}{l}{1s2s2p$^{2}$3d $^{6}$F$_{J}$ - 1s2p$^{3}$3d $^{6}$D$%
_{J^{\prime }}$} &  &  &  &  &  &  &  &  &  &  &  \\
 &  &  &  & \multicolumn{1}{r}{$\pm 106$} &  &  & \multicolumn{1}{r}{$\pm 167$} &  &  & \multicolumn{1}{r}{$\pm 385$}
&  &  & \multicolumn{1}{r}{$\pm 783$} &  \\
1/2-* & 665.38 & 150290 & 665.83 & 150188 & \multicolumn{1}{r}{-102} & 666.07 & 150134 & \multicolumn{1}{r}{-156} & 663.74 & 150661 & \multicolumn{1}{r}{371} & 661.98 & 151062 & \multicolumn{1}{r}{772} \\
3/2-* & 665.65 & 150229 & 666.08 & 150132 & \multicolumn{1}{r}{-97} & 666.25 & 150094 & \multicolumn{1}{r}{-135} & 663.93 & 150618 & \multicolumn{1}{r}{389} & 662.18 & 151016 & \multicolumn{1}{r}{787} \\
5/2-* & 665.92 & 150168 & 666.32 & 150078 & \multicolumn{1}{r}{-90} & 666.56 & 150024 & \multicolumn{1}{r}{-144} & 664.25 & 150546 & \multicolumn{1}{r}{378} & 662.49 & 150946 & \multicolumn{1}{r}{777} \\
7/2-* & 666.27 & 150089 & 666.69 & 149995 & \multicolumn{1}{r}{-95} & 666.74 & 149984 & \multicolumn{1}{r}{-106} & 664.65 & 150455 & \multicolumn{1}{r}{366} & 662.88 & 150857 & \multicolumn{1}{r}{768} \\
9/2-* & 666.63 & 150008 & 667.10 & 149903 & \multicolumn{1}{r}{-106} & 667.35 & 149846 & \multicolumn{1}{r}{-162} & 665.08 & 150358 & \multicolumn{1}{r}{350} & 663.31 & 150759 & \multicolumn{1}{r}{751} \\
11/-9/2 & 666.99 & 149927 & 667.47 & 149819 & \multicolumn{1}{r}{-108} & 667.73 & 149761 & \multicolumn{1}{r}{-166} & 665.51 & 150261 & \multicolumn{1}{r}{333} & 663.72 & 150666 & \multicolumn{1}{r}{739} \\
AV & 666.41 & 150058 & 666.86 & 149957 & \multicolumn{1}{r}{-101} & 667.06 & 149911 & \multicolumn{1}{r}{-147} & 664.83 & 150415 & \multicolumn{1}{r}{357} & 663.06 & 150817 & \multicolumn{1}{r}{759} \\
\multicolumn{1}{l}{QED} &  &  &  & \multicolumn{1}{r}{-23.2} &  &
& \multicolumn{1}{r}{-23.2} &  &  &  &  &  &  &
\\
\multicolumn{1}{l}{HO} &  &  &  & \multicolumn{1}{r}{270.8} &  &  & \multicolumn{1}{r}{84.6} &  &  &  &  &  &  &  \\
\multicolumn{1}{l}{AV$^{T}$} &  &  & 665.76 & 150205 &
\multicolumn{1}{r}{147} & 666.79 &
149972 & \multicolumn{1}{r}{-86} &  &  &  &  &  &  \\
\multicolumn{1}{l}{nonrel} &  &  &  &  &  & 670.72 & 149094 &
\multicolumn{1}{r}{-964} &
 &  &  &  &  &  \\
\multicolumn{4}{l}{1s2s2p$^{2}$3d $^{6}$D$_{J}$ - 1s2p$^{3}$3d $^{6}$D$%
_{J^{\prime }}$} &  &  &  &  &  &  &  &  &  &  &  \\
&  &  &  & $\pm 106$ &  &  & $\pm 167$ &  &  & $\pm 385$
&  &  & $\pm 783$ &  \\
1/2-* & 686.42 & 145683 & 685.35 & 145911 & \multicolumn{1}{r}{  227} & 688.18 & 145311 & \multicolumn{1}{r}{-373} & 684.20 & 146156 & \multicolumn{1}{r}{473} & 682.24 & 146576 & \multicolumn{1}{r}{893} \\
3/2-* & 686.39 & 145690 & 685.32 & 145917 & \multicolumn{1}{r}{  227} & 688.15 & 145317 & \multicolumn{1}{r}{-373} & 684.18 & 146160 & \multicolumn{1}{r}{471} & 682.22 & 146580 & \multicolumn{1}{r}{891} \\
5/2-* & 686.39 & 145690 & 685.30 & 145921 & \multicolumn{1}{r}{  232} & 688.15 & 145317 & \multicolumn{1}{r}{-373} & 684.47 & 146098 & \multicolumn{1}{r}{409} & 682.23 & 146578 & \multicolumn{1}{r}{888} \\
7/2-* & 686.51 & 145664 & 685.41 & 145898 & \multicolumn{1}{r}{  234} & 688.27 & 145292 & \multicolumn{1}{r}{-372} & 684.27 & 146141 & \multicolumn{1}{r}{477} & 682.35 & 146552 & \multicolumn{1}{r}{888} \\
9/2-7/2 & 686.87 & 145588 & 685.76 & 145824 & \multicolumn{1}{r}{  236} & 688.63 & 145216 & \multicolumn{1}{r}{-372} & 684.71 & 146047 & \multicolumn{1}{r}{459} & 682.73 & 146471 & \multicolumn{1}{r}{883} \\
AV & 686.58 & 145649 & 685.49 & 145881 & \multicolumn{1}{r}{  233} & 688.34 & 145276 & \multicolumn{1}{r}{-372} & 684.44 & 146105 & \multicolumn{1}{r}{456} & 682.43 & 146536 & \multicolumn{1}{r}{887} \\
QED &  &  &  & \multicolumn{1}{r}{-23.1} & \multicolumn{1}{r}{} &
\multicolumn{1}{r}{} & \multicolumn{1}{r}{-23.0} &
\multicolumn{1}{r}{} & \multicolumn{1}{r}{} & \multicolumn{1}{r}{}
& \multicolumn{1}{r}{} & \multicolumn{1}{r}{} &
\multicolumn{1}{r}{} & \multicolumn{1}{r}{} \\
HO &  &  &  & 93.1 & \multicolumn{1}{r}{} & \multicolumn{1}{r}{} &
\multicolumn{1}{r}{209.8} & \multicolumn{1}{r}{} &
\multicolumn{1}{r}{} & \multicolumn{1}{r}{} & \multicolumn{1}{r}{}
& \multicolumn{1}{r}{} &
\multicolumn{1}{r}{} & \multicolumn{1}{r}{} \\
AV$^{T}$ &  &  & 685.15 & 145951 & \multicolumn{1}{r}{302} &
\multicolumn{1}{r}{687.46} & \multicolumn{1}{r}{145463} &
\multicolumn{1}{r}{ -186} & \multicolumn{1}{r}{} &
\multicolumn{1}{r}{} & \multicolumn{1}{r}{} &
\multicolumn{1}{r}{} & \multicolumn{1}{r}{} & \multicolumn{1}{r}{} \\
N-REL &  &  &  &  & \multicolumn{1}{r}{} &
\multicolumn{1}{r}{692.20} & \multicolumn{1}{r}{144467} &
\multicolumn{1}{r}{-1182} & \multicolumn{1}{r}{} &
\multicolumn{1}{r}{} & \multicolumn{1}{r}{} & \multicolumn{1}{r}{}
& \multicolumn{1}{r}{} & \multicolumn{1}{r}{}
\\
\multicolumn{5}{l}{1s2s2p$^{2}$3d $^{6}$P$_{J}$ - 1s2p$^{3}$3d $^{6}$D$%
_{J^{\prime }}$} &  &  &  &  &  &  &  &  &  &  \\
&  &  &  & \multicolumn{1}{r}{$\pm 869$} & \multicolumn{1}{r}{} &
\multicolumn{1}{r}{} & \multicolumn{1}{r}{$\pm 650$} &
\multicolumn{1}{r}{} & \multicolumn{1}{r}{} &
\multicolumn{1}{r}{$\pm 1705$} & \multicolumn{1}{r}{} &
\multicolumn{1}{r}{
} & \multicolumn{1}{r}{$\pm 590$} & \multicolumn{1}{r}{} \\
3/2-* & 698.06 & 143254 & 693.88 & 144117 & \multicolumn{1}{r}{  863} & 701.21 & 142611 & \multicolumn{1}{r}{-644} & 689.94 & 144940 & \multicolumn{1}{r}{1686} & 695.23 & 143837 & \multicolumn{1}{r}{583} \\
5/2-* & 698.57 & 143150 & 694.42 & 144005 & \multicolumn{1}{r}{  855} & 701.76 & 142499 & \multicolumn{1}{r}{-651} & 690.42 & 144839 & \multicolumn{1}{r}{1690} & 695.76 & 143728 & \multicolumn{1}{r}{578} \\
7/2-* & 698.95 & 143072 & 694.73 & 143941 & \multicolumn{1}{r}{  869} & 702.08 & 142434 & \multicolumn{1}{r}{-638} & 690.71 & 144779 & \multicolumn{1}{r}{1707} & 696.08 & 143662 & \multicolumn{1}{r}{590} \\
AV & 698.63 & 143138 & 694.25 & 144041 & \multicolumn{1}{r}{  902} & 701.59 & 142534 & \multicolumn{1}{r}{-604} & 690.27 & 144871 & \multicolumn{1}{r}{1732} & 695.60 & 143762 & \multicolumn{1}{r}{624} \\
QED &  &  &  & \multicolumn{1}{r}{-23.0} & \multicolumn{1}{r}{} &
\multicolumn{1}{r}{} & \multicolumn{1}{r}{-22.9} &
\multicolumn{1}{r}{} & \multicolumn{1}{r}{} & \multicolumn{1}{r}{}
& \multicolumn{1}{r}{} & \multicolumn{1}{r}{} &
\multicolumn{1}{r}{} & \multicolumn{1}{r}{} \\
HO &  &  &  & 282.0 & \multicolumn{1}{r}{} & \multicolumn{1}{r}{}
& \multicolumn{1}{r}{112.1} & \multicolumn{1}{r}{} &
\multicolumn{1}{r}{} & \multicolumn{1}{r}{} & \multicolumn{1}{r}{}
& \multicolumn{1}{r}{} &
\multicolumn{1}{r}{} & \multicolumn{1}{r}{} \\
AV$^{T}$ &  &  & 693.00 & 144300 & \multicolumn{1}{r}{1162} &
\multicolumn{1}{r}{701.11} & \multicolumn{1}{r}{142623} &
\multicolumn{1}{r}{ -515} & \multicolumn{1}{r}{} &
\multicolumn{1}{r}{} & \multicolumn{1}{r}{} &
\multicolumn{1}{r}{} & \multicolumn{1}{r}{} & \multicolumn{1}{r}{} \\
N-REL &  &  &  &  & \multicolumn{1}{r}{} &
\multicolumn{1}{r}{705.59} & \multicolumn{1}{r}{141725} &
\multicolumn{1}{r}{-1413} & \multicolumn{1}{r}{} &
\multicolumn{1}{r}{} & \multicolumn{1}{r}{} & \multicolumn{1}{r}{}
& \multicolumn{1}{r}{} & \multicolumn{1}{r}{}
\end{tabular}
\end{ruledtabular}
\end{table*}
\end{center}

Similarly, after studying details of transitions theoretically and
experimentally described above, and comparing with
multi-configuration Hartree-Fock (MCHF) and multi-configuration
Dirac-Fock (MCDF) calculations of O IV by this work, we were able
to assign these unidentified observed lines as the 1s2s2p$^{2}$3d
$^{6}$L-1s2p$^{3}$3d $^{6}$D$^{o}$, L=F, D and P electric-dipole
transitions in O IV. Results of the identification and
measurements of wavelengths of the transitions are listed in Table
I. Errors of measured wavelengths of $\pm $0.06 \AA\ are small
mainly from calibration and curve fitting. The latter includes
experimental and statistical errors. In Table I average
theoretical transition energy AV is the center of gravity of the
1s2s2p$^{2}$3d $^{6}$L-1s2p$^{3}$3d $^{6}$D$^{o}$ transition
energies (computed from fine structure lines calculated by this
work) with results of theoretical analysis. Experimental
transition energy AV is the center of gravity of the
1s2s2p$^{2}$3d $^{6}$L-1s2p$^{3}$3d $^{6}$D$^{o}$ transition
energies (computed from observed lines) with results of
experimental transition rate analysis. AV$^{T}$ is summation of
above average transition energy (AV), QED effect (QED) and
higher-order correction (HO). Errors for calculated transition
energies in Table I are the root mean squared differences of
calculated and experimental transition energies as given below in
the table. We also list calculated non-relativistic transition
energies (N-REL) by SCHF method. In Table I we present measured
fine structure wavelength values and theoretical values for O IV.
The measured and calculated results are consistent after
considering experimental and theoretical errors. Fourteen lines
are new observations with wavelength accuracy of $\pm $%
0.06 \AA .

\begin{table}
\caption{\label{tab:table1}Wavelengths $\lambda $ (in \AA ) for the 1s$^{2}$2p3p $^{1}$D$%
_{2}$-1s$^{2}$2p4d $^{1}$F$_{3}$ transitions in O V.}
\begin{ruledtabular}
\begin{tabular}{cccc}
$\lambda obs^{a}$ & $\lambda obs^{b}$ & $\lambda scdf^{a}$ &
$\lambda schf^{a}$ \\   \hline
$\pm 0.06$ & $\pm 0.03$ &  &  \\
666.27 & 666.819 & 651.12 &  \\
599.85 &  &  & 589.09
\end{tabular}
\end{ruledtabular}
a this work. b Elden~\cite{o3}.
\end{table}

It is noted that in Fig. 2 the line for the 1s2s2p$^{2}$3d $^{6}$F$_{7/2}$-1s2p$^{3}$3d $%
^{6}$D$_{J^{\prime =\ast }}^{o}$ transition in O IV is much stronger than
the SCHF result. The line is a blend of the 1s2s2p$^{2}$3d $^{6}$F$_{J=7/2}$-1s2p$^{3}$3d $%
^{6}$D$_{J^{\prime =\ast }}^{o}$ transition in O IV and a line at
the wavelength of 666.27$\pm $0.06 \AA . The latter was identified
as the 1s$^{2}$2p3p $^{1}$D$_{2}$-1s$^{2}$2p4d $^{1}$F$_{3}$
transition at the wavelength of 666.819$\pm $0.03 \AA
~\cite{o1,o2,o3}. In this work we could resolve the set of lines
and improve the wavelength accuracy of the 1s$^{2}$2p3p
$^{1}$D$_{2}$-1s$^{2}$2p4d $^{1}$F$_{3}$ transition to 666.27$\pm
$0.06 \AA . We list results in Table II.
For the 1s$^{2}$2p3p $^{1}$D$_{2}$ state there are two couplings, 1s$%
^{2}$2p(j=1/2)3p(j'=3/2) $^{1}$D$_{2}$ and 1s$^{2}$2p(j=3/2)3p(j'=1/2) $^{1}$%
D$_{2}$. For the 1s$^{2}$2p4d $^{1}$F$_{3}$ state there are two
couplings, 1s$^{2}$2p(j=1/2)4d(j'=5/2) $^{1}$F$_{3}$ and 1s$^{2}$%
2p(j=3/2)4d(j'=3/2) $^{1}$F$_{3}$. Their energies are different.
MCHF and MCDF methods handle it in different ways and give
different results. MCHF calculation gives the minimum energies
among both configurations. In Table II are listed wavelengths of
SCHF and SCDF calculations. We studied the
spectrum in the wavelength range around 590 \AA\ %
and found a strong unidentified line located at 599.85$\pm $0.06
\AA . After studying its details, we assign it as the 1s$^{2}$2p3p
$^{1}$D$_{2}$-1s$^{2}$2p4d $^{1}$F$_{3}$ transition in O V and
list it in Table II.

\begin{center}
\begin{table*}
\caption{\label{tab:table1}Energies E (in cm-1) and wavelengths
$\lambda $ (in \AA ) for the 1s2s2p$^{2}$3d
$^{6}$L$_{J}$-1s2p$^{3}$3d $^{6}$D$_{J\prime }$, L= F, D, P
transitions in F V by this work. We list differences dE between
theoretical and experimental transition energies for the
transitions.}
\begin{ruledtabular}
\begin{tabular}{llrlrrlrrlrrlrr}
J-J' & $\lambda $exp & Eexp & $\lambda $mchf & Emchf & dEmchf &
$\lambda $schf & Eschf & dEschf & $\lambda $mcdf
& Emcdf & dEmcdf & $\lambda $scdf & Escdf & dEscdf \\
  \hline
 & $\pm $0.10 &  &  &  &  &  &  &  &  &  &  &  &  &  \\
\multicolumn{5}{l}{1s2s2p$^{2}$3d $^{6}$F$_{J}$ - 1s2p$^{3}$3d $^{6}$D$%
_{J^{\prime }}$} &  &  &  &  &  &  &  &  &  &  \\
&  &  &  & $\pm 249$ & \multicolumn{1}{r}{} & \multicolumn{1}{r}{}
& \multicolumn{1}{r}{$\pm 156$} & \multicolumn{1}{r}{} &
\multicolumn{1}{r}{} & \multicolumn{1}{r}{$\pm 596$} &
\multicolumn{1}{r}{} & \multicolumn{1}{r}{}
& \multicolumn{1}{r}{$\pm 1041$} & \multicolumn{1}{r}{} \\
1/2-* & 571.76 & 174899 & 571.40 & 175009 & \multicolumn{1}{r}{  110} & 572.27 & 174743 & \multicolumn{1}{r}{ -156} & 569.83 & 175491 & \multicolumn{1}{r}{  592} & 568.94 & 175765 & \multicolumn{1}{r}{  867} \\
3/2-* & 572.01 & 174822 & 571.64 & 174935 & \multicolumn{1}{r}{  113} & 572.52 & 174666 & \multicolumn{1}{r}{ -156} & 570.07 & 175417 & \multicolumn{1}{r}{  595} & 568.17 & 176004 & \multicolumn{1}{r}{ 1182} \\
5/2-* & 572.41 & 174700 & 572.05 & 174810 & \multicolumn{1}{r}{  110} & 572.92 & 174544 & \multicolumn{1}{r}{ -156} & 570.48 & 175291 & \multicolumn{1}{r}{  591} & 569.59 & 175565 & \multicolumn{1}{r}{  865} \\
7/2-* & 572.91 & 174547 & 572.55 & 174657 & \multicolumn{1}{r}{  110} & 573.42 & 174392 & \multicolumn{1}{r}{ -155} & 571.02 & 175125 & \multicolumn{1}{r}{  578} & 570.13 & 175399 & \multicolumn{1}{r}{  851} \\
9/2-* & 573.48 & 174374 & 572.58 & 174648 & \multicolumn{1}{r}{  274} & 573.99 & 174219 & \multicolumn{1}{r}{ -155} & 571.62 & 174941 & \multicolumn{1}{r}{  567} & 570.75 & 175208 & \multicolumn{1}{r}{  834} \\
11/2-9/2 & 574.05 & 174201 & 573.15 & 174474 & \multicolumn{1}{r}{  274} & 574.56 & 174046 & \multicolumn{1}{r}{ -155} & 572.23 & 174755 & \multicolumn{1}{r}{  554} & 571.37 & 175018 & \multicolumn{1}{r}{  817} \\
AV & 573.16 & 174471 & 572.52 & 174668 & \multicolumn{1}{r}{  196} & 573.67 & 174316 & \multicolumn{1}{r}{ -155} & 571.28 & 175044 & \multicolumn{1}{r}{  573} & 570.31 & 175343 & \multicolumn{1}{r}{  871} \\
QED &  &  &  & -41.1 & \multicolumn{1}{r}{} & \multicolumn{1}{r}{}
& \multicolumn{1}{r}{-41.1} & \multicolumn{1}{r}{} &
\multicolumn{1}{r}{} & \multicolumn{1}{r}{} & \multicolumn{1}{r}{}
& \multicolumn{1}{r}{} &
\multicolumn{1}{r}{} & \multicolumn{1}{r}{} \\
HO &  &  &  & 104.0 & \multicolumn{1}{r}{} & \multicolumn{1}{r}{}
& \multicolumn{1}{r}{121.1} & \multicolumn{1}{r}{} &
\multicolumn{1}{r}{} & \multicolumn{1}{r}{} & \multicolumn{1}{r}{}
& \multicolumn{1}{r}{} &
\multicolumn{1}{r}{} & \multicolumn{1}{r}{} \\
AV$^{T}$ &  &  & 572.31 & 174731 & \multicolumn{1}{r}{260} &
\multicolumn{1}{r}{573.41} & \multicolumn{1}{r}{174396} &
\multicolumn{1}{r}{ -75} & \multicolumn{1}{r}{} &
\multicolumn{1}{r}{} & \multicolumn{1}{r}{} &
\multicolumn{1}{r}{} & \multicolumn{1}{r}{} & \multicolumn{1}{r}{} \\
N-REL &  &  &  &  & \multicolumn{1}{r}{} &
\multicolumn{1}{r}{578.11} & \multicolumn{1}{r}{172977} &
\multicolumn{1}{r}{-1494} & \multicolumn{1}{r}{} &
\multicolumn{1}{r}{} & \multicolumn{1}{r}{} & \multicolumn{1}{r}{}
& \multicolumn{1}{r}{} & \multicolumn{1}{r}{}
\\
\multicolumn{5}{l}{1s2s2p$^{2}$3d $^{6}$D$_{J}$ - 1s2p$^{3}$3d $^{6}$D$%
_{J^{\prime }}$} &  &  &  &  &  &  &  &  &  &  \\
&  &  &  & $\pm $104  & \multicolumn{1}{r}{} &
\multicolumn{1}{r}{} & \multicolumn{1}{r}{$\pm $564} &
\multicolumn{1}{r}{} & \multicolumn{1}{r}{} &
\multicolumn{1}{r}{$\pm $549} & \multicolumn{1}{r}{} &
\multicolumn{1}{r}{}
& \multicolumn{1}{r}{$\pm $830} & \multicolumn{1}{r}{} \\
1/2-* & 591.85 & 168962 & 591.51 & 169059 & \multicolumn{1}{r}{   97} & 593.83 & 168398 & \multicolumn{1}{r}{ -563} & 589.97 & 169500 & \multicolumn{1}{r}{  538} & 588.97 & 169788 & \multicolumn{1}{r}{  826} \\
3/2-* & 591.82 & 168970 & 591.48 & 169067 & \multicolumn{1}{r}{   97} & 593.80 & 168407 & \multicolumn{1}{r}{ -563} & 589.93 & 169512 & \multicolumn{1}{r}{  541} & 588.94 & 169797 & \multicolumn{1}{r}{  826} \\
5/2-* & 591.86 & 168959 & 591.51 & 169059 & \multicolumn{1}{r}{  100} & 593.84 & 168396 & \multicolumn{1}{r}{ -563} & 589.96 & 169503 & \multicolumn{1}{r}{  544} & 588.98 & 169785 & \multicolumn{1}{r}{  826} \\
7/2-* & 592.05 & 168905 & 591.69 & 169007 & \multicolumn{1}{r}{  103} & 594.03 & 168342 & \multicolumn{1}{r}{ -563} & 590.14 & 169451 & \multicolumn{1}{r}{  547} & 589.18 & 169727 & \multicolumn{1}{r}{  823} \\
9/2-* & 592.52 & 168771 & 592.15 & 168876 & \multicolumn{1}{r}{  105} & 594.50 & 168209 & \multicolumn{1}{r}{ -562} & 590.70 & 169291 & \multicolumn{1}{r}{  520} & 589.70 & 169578 & \multicolumn{1}{r}{  807} \\
AV & 592.12 & 168883 & 591.77 & 168985 & \multicolumn{1}{r}{  102} & 594.10 & 168321 & \multicolumn{1}{r}{ -563} & 590.25 & 169419 & \multicolumn{1}{r}{  536} & 589.27 & 169702 & \multicolumn{1}{r}{  819} \\
QED &  &  &  & -40.8 & \multicolumn{1}{r}{} & \multicolumn{1}{r}{}
& \multicolumn{1}{r}{-40.8} & \multicolumn{1}{r}{} &
\multicolumn{1}{r}{} & \multicolumn{1}{r}{} & \multicolumn{1}{r}{}
& \multicolumn{1}{r}{} &
\multicolumn{1}{r}{} & \multicolumn{1}{r}{} \\
HO &  &  &  & 249.7 & \multicolumn{1}{r}{} & \multicolumn{1}{r}{}
& \multicolumn{1}{r}{63.4} & \multicolumn{1}{r}{} &
\multicolumn{1}{r}{} & \multicolumn{1}{r}{} & \multicolumn{1}{r}{}
& \multicolumn{1}{r}{} &
\multicolumn{1}{r}{} & \multicolumn{1}{r}{} \\
AV$^{T}$ &  &  & 591.04 & 169194 & \multicolumn{1}{r}{311} &
\multicolumn{1}{r}{594.02} & \multicolumn{1}{r}{168344} &
\multicolumn{1}{r}{ -539} & \multicolumn{1}{r}{} &
\multicolumn{1}{r}{} & \multicolumn{1}{r}{} &
\multicolumn{1}{r}{} & \multicolumn{1}{r}{} & \multicolumn{1}{r}{} \\
N-REL &  &  &  &  & \multicolumn{1}{r}{} &
\multicolumn{1}{r}{598.88} & \multicolumn{1}{r}{166978} &
\multicolumn{1}{r}{-1905} & \multicolumn{1}{r}{} &
\multicolumn{1}{r}{} & \multicolumn{1}{r}{} & \multicolumn{1}{r}{}
& \multicolumn{1}{r}{} & \multicolumn{1}{r}{}
\\
\multicolumn{5}{l}{1s2s2p$^{2}$3d $^{6}$P$_{J}$ - 1s2p$^{3}$3d $^{6}$D$%
_{J^{\prime }}$} &  &  &  &  &  &  &  &  &  &  \\
&  &  &  & $\pm $353 & \multicolumn{1}{r}{} & \multicolumn{1}{r}{}
& \multicolumn{1}{r}{$\pm $1861} & \multicolumn{1}{r}{} &
\multicolumn{1}{r}{} & \multicolumn{1}{r}{$\pm $1235} &
\multicolumn{1}{r}{} & \multicolumn{1}{r}{
} & \multicolumn{1}{r}{$\pm $475} & \multicolumn{1}{r}{} \\
3/2-* & 604.10 & 165536 & 603.27 & 165763 & \multicolumn{1}{r}{  228} & 610.97 & 163674 & \multicolumn{1}{r}{-1861} & 599.75 & 166736 & \multicolumn{1}{r}{ 1201} & 605.84 & 165060 & \multicolumn{1}{r}{ -475} \\
5/2-* & 604.88 & 165322 & 603.99 & 165566 & \multicolumn{1}{r}{  244} & 611.75 & 163465 & \multicolumn{1}{r}{-1857} & 600.45 & 166542 & \multicolumn{1}{r}{ 1220} & 606.60 & 164853 & \multicolumn{1}{r}{ -469} \\
7/2-* & 605.36 & 165191 & 604.03 & 165555 & \multicolumn{1}{r}{  364} & 612.22 & 163340 & \multicolumn{1}{r}{-1851} & 600.87 & 166425 & \multicolumn{1}{r}{ 1234} & 607.08 & 164723 & \multicolumn{1}{r}{ -468} \\
AV & 604.64 & 165388 & 603.68 & 165651 & \multicolumn{1}{r}{  263} & 611.51 & 163530 & \multicolumn{1}{r}{-1857} & 600.23 & 166602 & \multicolumn{1}{r}{ 1215} & 606.37 & 164916 & \multicolumn{1}{r}{ -472} \\
QED &  &  &  &  -40.6  & \multicolumn{1}{r}{} &
\multicolumn{1}{r}{} & \multicolumn{1}{r}{-40.5 } &
\multicolumn{1}{r}{} & \multicolumn{1}{r}{} & \multicolumn{1}{r}{}
& \multicolumn{1}{r}{} & \multicolumn{1}{r}{} &
\multicolumn{1}{r}{} & \multicolumn{1}{r}{} \\
HO &  &  &  & 53.0 & \multicolumn{1}{r}{} & \multicolumn{1}{r}{} &
\multicolumn{1}{r}{175.8} & \multicolumn{1}{r}{} &
\multicolumn{1}{r}{} & \multicolumn{1}{r}{} & \multicolumn{1}{r}{}
& \multicolumn{1}{r}{} &
\multicolumn{1}{r}{} & \multicolumn{1}{r}{} \\
AV$^{T}$ &  &  & 603.63 & 165663 & \multicolumn{1}{r}{275} &
\multicolumn{1}{r}{611.00} & \multicolumn{1}{r}{163665} &
\multicolumn{1}{r}{ -1723} & \multicolumn{1}{r}{} &
\multicolumn{1}{r}{} & \multicolumn{1}{r}{}
& \multicolumn{1}{r}{} & \multicolumn{1}{r}{} & \multicolumn{1}{r}{} \\
N-REL &  &  &  &  & \multicolumn{1}{r}{} &
\multicolumn{1}{r}{616.56} & \multicolumn{1}{r}{162190} &
\multicolumn{1}{r}{-3197} & \multicolumn{1}{r}{} &
\multicolumn{1}{r}{} & \multicolumn{1}{r}{} & \multicolumn{1}{r}{}
& \multicolumn{1}{r}{} & \multicolumn{1}{r}{}
\end{tabular}
\end{ruledtabular}
\end{table*}
\end{center}

We obtained spectra at a $^{20}$(HF)$^{+}$ beam energy of 2.5 MeV
for fluorine. Through the use of optical refocusing we achieved
spectroscopic line width of 0.7 \AA . Using similar experimental
analysis as described above we obtained the wavelength accuracy of
$\pm $0.10 \AA\ for the 1s2s2p$^{2}$3d $^{6}$L-1s2p$^{3}$3d
$^{6}$D transitions in the wavelength region of 570-620 \AA
~\cite{f1,f2}. In Table III all observed
lines for the 1s2s2p$^{2}$3d $^{6}$L - 1s2p$^{3}$%
3d $^{6}$D$^{o}$ , L=F, D, P transitions in F V are reported.
Fourteen lines are new observations. The strongest fine structure
component is the 1s2s2p$^{2}$3d $^{6}$F$_{11/2}$-1s2p$^{3}$3d $^{6}$D$%
_{9/2}^{o}$ transition at the wavelength of 574.05$\pm $0.10 \AA .

\begin{center}
\begin{table*}
\caption{\label{tab:table1}Energies E (in cm-1) and wavelengths $\lambda $ (in \AA ) for the 1s2s2p%
$^{2}$3d $^{6}$L$_{J}$-1s2p$^{3}$3d $^{6}$D$_{J\prime }^{o}$, L=
F, D, P transitions in Ne VI by this work. We list differences dE
between theoretical and experimental transition energies for the
transitions.}
\begin{ruledtabular}
\begin{tabular}{llrlrrlrrlrrlrr}
J-J' & $\lambda $exp & Eexp & $\lambda $mchf & Emchf & dEmchf &
$\lambda $schf & Eschf & dEschf & $\lambda $mcdf
& Emcdf & dEmcdf & $\lambda $scdf & Escdf & dEscdf \\
  \hline
 & $\pm $0.05 &  &  &  &  &  &  &  &  &  &  &  &  &  \\
\multicolumn{5}{l}{1s2s2p$^{2}$3d $^{6}$F$_{J}$ - 1s2p$^{3}$3d $^{6}$D$%
_{J^{\prime }}$} &  &  &  &  &  &  &  &  &  &  \\
&  &  &  & $\pm $3448 & \multicolumn{1}{r}{} &
\multicolumn{1}{r}{} & \multicolumn{1}{r}{$\pm $180} &
\multicolumn{1}{r}{} & \multicolumn{1}{r}{} &
\multicolumn{1}{r}{$\pm $811} & \multicolumn{1}{r}{} &
\multicolumn{1}{r}{}
& \multicolumn{1}{r}{$\pm $946} & \multicolumn{1}{r}{} \\
1/2-* & 500.40 & 199840 & 492.42 & 203079 & \multicolumn{1}{r}{ 3239} & 500.79 & 199684 & \multicolumn{1}{r}{ -156} & 498.37 & 200654 & \multicolumn{1}{r}{  814} & 498.03 & 200791 & \multicolumn{1}{r}{  951} \\
3/2-* & 500.66 & 199736 & 492.69 & 202967 & \multicolumn{1}{r}{ 3231} & 501.12 & 199553 & \multicolumn{1}{r}{ -183} & 498.67 & 200533 & \multicolumn{1}{r}{  797} & 498.35 & 200662 & \multicolumn{1}{r}{  926} \\
5/2-* & 501.18 & 199529 & 493.12 & 202790 & \multicolumn{1}{r}{ 3261} & 501.62 & 199354 & \multicolumn{1}{r}{ -175} & 499.18 & 200329 & \multicolumn{1}{r}{  799} & 498.86 & 200457 & \multicolumn{1}{r}{  928} \\
7/2-* & 501.84 & 199267 & 493.69 & 202556 & \multicolumn{1}{r}{ 3290} & 502.27 & 199096 & \multicolumn{1}{r}{ -171} & 499.88 & 200048 & \multicolumn{1}{r}{  781} & 499.54 & 200184 & \multicolumn{1}{r}{  917} \\
9/2-* & 502.71 & 198922 & 494.26 & 202323 & \multicolumn{1}{r}{ 3401} & 503.03 & 198795 & \multicolumn{1}{r}{ -127} & 500.69 & 199724 & \multicolumn{1}{r}{  803} & 500.33 & 199868 & \multicolumn{1}{r}{  946} \\
11/2-9/2 & 503.45 & 198629 & 496.86 & 201264 & \multicolumn{1}{r}{ 2634} & 503.80 & 198491 & \multicolumn{1}{r}{ -138} & 501.53 & 199390 & \multicolumn{1}{r}{  760} & 501.17 & 199533 & \multicolumn{1}{r}{  904} \\
AV & 502.23 & 199111 & 494.49 & 202227 & \multicolumn{1}{r}{ 3116} & 502.62 & 198959 & \multicolumn{1}{r}{ -152} & 500.26 & 199897 & \multicolumn{1}{r}{  786} & 499.91 & 200035 & \multicolumn{1}{r}{  924} \\
QED &  &  &  & -67.2 & \multicolumn{1}{r}{} & \multicolumn{1}{r}{}
& \multicolumn{1}{r}{-67.1} & \multicolumn{1}{r}{} &
\multicolumn{1}{r}{} & \multicolumn{1}{r}{} & \multicolumn{1}{r}{}
& \multicolumn{1}{r}{} &
\multicolumn{1}{r}{} & \multicolumn{1}{r}{} \\
HO &  &  &  & 94.1 & \multicolumn{1}{r}{} & \multicolumn{1}{r}{} &
\multicolumn{1}{r}{-2.3} & \multicolumn{1}{r}{} &
\multicolumn{1}{r}{} & \multicolumn{1}{r}{} & \multicolumn{1}{r}{}
& \multicolumn{1}{r}{} &
\multicolumn{1}{r}{} & \multicolumn{1}{r}{} \\
AV$^{T}$ &  &  & 494.43 & 202254 & \multicolumn{1}{r}{3143} &
\multicolumn{1}{r}{502.79} & \multicolumn{1}{r}{198890} &
\multicolumn{1}{r}{ -221} & \multicolumn{1}{r}{} &
\multicolumn{1}{r}{} & \multicolumn{1}{r}{} &
\multicolumn{1}{r}{} & \multicolumn{1}{r}{} & \multicolumn{1}{r}{} \\
N-REL &  &  &  &  & \multicolumn{1}{r}{} &
\multicolumn{1}{r}{508.95} & \multicolumn{1}{r}{196483} &
\multicolumn{1}{r}{-2628} & \multicolumn{1}{r}{} &
\multicolumn{1}{r}{} & \multicolumn{1}{r}{} & \multicolumn{1}{r}{}
& \multicolumn{1}{r}{} & \multicolumn{1}{r}{}
\\
\multicolumn{5}{l}{1s2s2p$^{2}$3d $^{6}$D$_{J}$ - 1s2p$^{3}$3d $^{6}$D$%
_{J^{\prime }}$} &  &  &  &  &  &  &  &  &  &  \\
&  &  &  & $\pm $2509 & \multicolumn{1}{r}{} &
\multicolumn{1}{r}{} & \multicolumn{1}{r}{$\pm $671} &
\multicolumn{1}{r}{} & \multicolumn{1}{r}{} &
\multicolumn{1}{r}{$\pm $745} & \multicolumn{1}{r}{} &
\multicolumn{1}{r}{}
& \multicolumn{1}{r}{$\pm $891} & \multicolumn{1}{r}{} \\
1/2-* & 519.65 & 192437 & 513.56 & 194719 & \multicolumn{1}{r}{ 2282} & 521.42 & 191784 & \multicolumn{1}{r}{ -653} & 517.81 & 193121 & \multicolumn{1}{r}{  684} & 517.41 & 193270 & \multicolumn{1}{r}{  833} \\
3/2-* & 519.63 & 192445 & 513.55 & 194723 & \multicolumn{1}{r}{ 2278} & 521.41 & 191788 & \multicolumn{1}{r}{ -657} & 517.78 & 193132 & \multicolumn{1}{r}{  688} & 517.40 & 193274 & \multicolumn{1}{r}{  829} \\
5/2-* & 519.70 & 192419 & 513.49 & 194746 & \multicolumn{1}{r}{ 2327} & 521.49 & 191758 & \multicolumn{1}{r}{ -660} & 517.83 & 193114 & \multicolumn{1}{r}{  695} & 517.48 & 193244 & \multicolumn{1}{r}{  825} \\
7/2-* & 520.14 & 192256 & 513.57 & 194715 & \multicolumn{1}{r}{ 2459} & 521.75 & 191663 & \multicolumn{1}{r}{ -593} & 518.10 & 193013 & \multicolumn{1}{r}{  757} & 517.74 & 193147 & \multicolumn{1}{r}{  891} \\
9/2-* & 520.81 & 192009 & 513.94 & 194575 & \multicolumn{1}{r}{ 2567} & 522.33 & 191450 & \multicolumn{1}{r}{ -559} & 518.82 & 192745 & \multicolumn{1}{r}{  736} & 518.39 & 192905 & \multicolumn{1}{r}{  896} \\
AV & 520.17 & 192243 & 513.67 & 194676 & \multicolumn{1}{r}{ 2433} & 521.82 & 191635 & \multicolumn{1}{r}{ -608} & 518.22 & 192967 & \multicolumn{1}{r}{  724} & 517.84 & 193111 & \multicolumn{1}{r}{  868} \\
QED &  &  &  & -66.7 & \multicolumn{1}{r}{} & \multicolumn{1}{r}{}
& \multicolumn{1}{r}{-66.6} & \multicolumn{1}{r}{} &
\multicolumn{1}{r}{} & \multicolumn{1}{r}{} & \multicolumn{1}{r}{}
& \multicolumn{1}{r}{} &
\multicolumn{1}{r}{} & \multicolumn{1}{r}{} \\
HO &  &  &  & -16.5 & \multicolumn{1}{r}{} & \multicolumn{1}{r}{}
& \multicolumn{1}{r}{-85.3} & \multicolumn{1}{r}{} &
\multicolumn{1}{r}{} & \multicolumn{1}{r}{} & \multicolumn{1}{r}{}
& \multicolumn{1}{r}{} &
\multicolumn{1}{r}{} & \multicolumn{1}{r}{} \\
AV$^{T}$ &  &  & 513.89 & 194593 & \multicolumn{1}{r}{2350} &
\multicolumn{1}{r}{522.24} & \multicolumn{1}{r}{191483} &
\multicolumn{1}{r}{ -760} & \multicolumn{1}{r}{} &
\multicolumn{1}{r}{} & \multicolumn{1}{r}{} &
\multicolumn{1}{r}{} & \multicolumn{1}{r}{} & \multicolumn{1}{r}{} \\
N-REL &  &  &  &  & \multicolumn{1}{r}{} &
\multicolumn{1}{r}{527.59} & \multicolumn{1}{r}{189541} &
\multicolumn{1}{r}{-2702} & \multicolumn{1}{r}{} &
\multicolumn{1}{r}{} & \multicolumn{1}{r}{} & \multicolumn{1}{r}{}
& \multicolumn{1}{r}{} & \multicolumn{1}{r}{}
\\
\multicolumn{5}{l}{1s2s2p$^{2}$3d $^{6}$P$_{J}$ - 1s2p$^{3}$3d $^{6}$D$%
_{J^{\prime }}$} &  &  &  &  &  &  &  &  &  &  \\
&  &  &  & $\pm $8011 & \multicolumn{1}{r}{} &
\multicolumn{1}{r}{} & \multicolumn{1}{r}{$\pm $1274} &
\multicolumn{1}{r}{} & \multicolumn{1}{r}{} &
\multicolumn{1}{r}{$\pm $2526} & \multicolumn{1}{r}{} &
\multicolumn{1}{r}{
} & \multicolumn{1}{r}{$\pm $273} & \multicolumn{1}{r}{} \\
3/2-* & 537.00 & 186220 & 517.23 & 193338 & \multicolumn{1}{r}{ 7118} & 540.69 & 184949 & \multicolumn{1}{r}{-1271} & 530.01 & 188676 & \multicolumn{1}{r}{ 2456} & 536.33 & 186452 & \multicolumn{1}{r}{  233} \\
5/2-* & 538.10 & 185839 & 518.34 & 192924 & \multicolumn{1}{r}{ 7085} & 541.74 & 184590 & \multicolumn{1}{r}{-1249} & 530.92 & 188352 & \multicolumn{1}{r}{ 2513} & 537.32 & 186109 & \multicolumn{1}{r}{  270} \\
7/2-* & 538.69 & 185636 & 529.08 & 189007 & \multicolumn{1}{r}{ 3372} & 542.39 & 184369 & \multicolumn{1}{r}{-1266} & 531.50 & 188147 & \multicolumn{1}{r}{ 2511} & 537.94 & 185894 & \multicolumn{1}{r}{  259} \\
AV & 537.74 & 185963 & 520.23 & 192221 & \multicolumn{1}{r}{ 6259} & 541.42 & 184700 & \multicolumn{1}{r}{-1262} & 530.64 & 188450 & \multicolumn{1}{r}{ 2487} & 537.02 & 186214 & \multicolumn{1}{r}{  251} \\
QED &  &  &  & -66.4 & \multicolumn{1}{r}{} & \multicolumn{1}{r}{}
& \multicolumn{1}{r}{-66.2} & \multicolumn{1}{r}{} &
\multicolumn{1}{r}{} & \multicolumn{1}{r}{} & \multicolumn{1}{r}{}
& \multicolumn{1}{r}{} &
\multicolumn{1}{r}{} & \multicolumn{1}{r}{} \\
HO &  &  &  & -100.4 & \multicolumn{1}{r}{} & \multicolumn{1}{r}{}
& \multicolumn{1}{r}{-219.8} & \multicolumn{1}{r}{} &
\multicolumn{1}{r}{} & \multicolumn{1}{r}{} & \multicolumn{1}{r}{}
& \multicolumn{1}{r}{} &
\multicolumn{1}{r}{} & \multicolumn{1}{r}{} \\
AV$^{T}$ &  &  & 520.69 & 192054 & \multicolumn{1}{r}{6091} &
\multicolumn{1}{r}{542.26} & \multicolumn{1}{r}{184414} &
\multicolumn{1}{r}{ -1549} & \multicolumn{1}{r}{} &
\multicolumn{1}{r}{} & \multicolumn{1}{r}{}
& \multicolumn{1}{r}{} & \multicolumn{1}{r}{} & \multicolumn{1}{r}{} \\
N-REL &  &  &  &  & \multicolumn{1}{r}{} &
\multicolumn{1}{r}{547.62} & \multicolumn{1}{r}{182608} &
\multicolumn{1}{r}{-3354} & \multicolumn{1}{r}{} &
\multicolumn{1}{r}{} & \multicolumn{1}{r}{} & \multicolumn{1}{r}{}
& \multicolumn{1}{r}{} & \multicolumn{1}{r}{}
\end{tabular}
\end{ruledtabular}
\end{table*}
\end{center}

We obtained spectra at a $^{20}$Ne$^{+}$ ion beam energy of 4.0
MeV for neon. Through the use of optical refocusing we achieved
spectroscopic line width of 0.4 \AA\ in the second order spectrum.
Similarly, we obtained wavelength
accuracy of $\pm $0.05 \AA\ for the 1s2s2p$^{2}$3d $^{6}$L - 1s2p$^{3}$3d $%
^{6}$D$^{o}$ transitions in the wavelength region of 490-550 \AA
~\cite{ne1,ne2}. In Table IV we present
measured fine structure wavelength values and theoretical values for the 1s2s2p$^{2}$3d $^{6}$L$_{J}$-1s2p$%
^{3}$3d $^{6}$D$_{J\prime }^{o}$, L=F, D, P transitions for Ne VI.
 Fourteen lines are new observations. The
strongest fine structure component is the 1s2s2p$^{2}$3d $%
^{6} $F$_{11/2}$-1s2p$^{3}$3d $^{6}$D$_{9/2}^{o}$ transition at
the wavelength of 503.45$\pm $0.05 \AA .

\begin{figure}[tbp]
\centerline{\includegraphics*[scale=1.00]{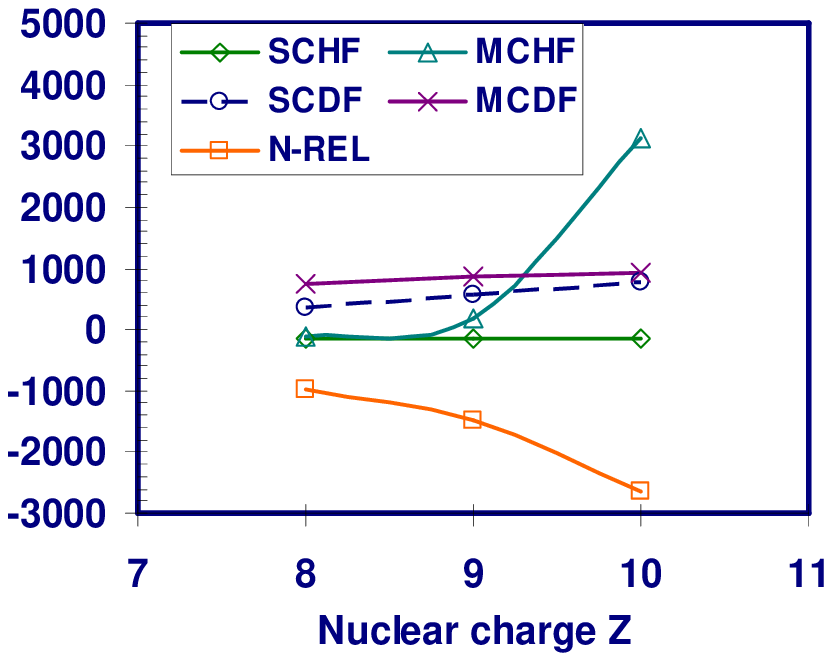}}
\caption{Difference (in cm$^{-1}$) between theoretical and
experimental transition energies for the 1s2s2p$^{2}$3d $^{6}$F -
1s2p$^{3}$3d $^{6}$D$^{o}$ transitions.} \label{fig4}
\end{figure}
\begin{figure}[tbp]
\centerline{\includegraphics*[scale=1.00]{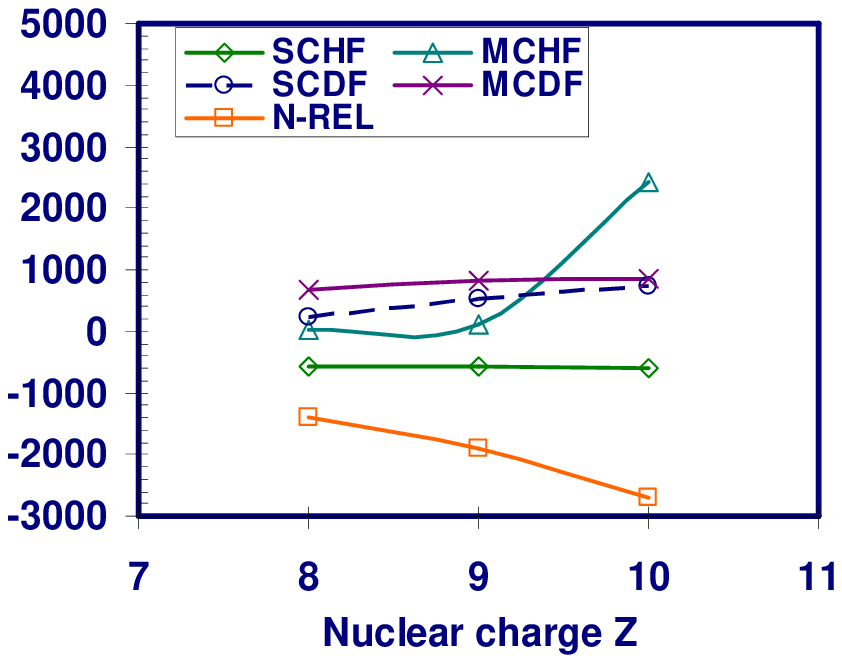}}
\caption{Difference (in cm$^{-1}$) between theoretical and
experimental transition energies for the 1s2s2p$^{2}$3d $^{6}$D -
1s2p$^{3}$3d $^{6}$D$^{o}$ transitions.} \label{fig5}
\end{figure}
\begin{figure}[tbp]
\centerline{\includegraphics*[scale=1.00]{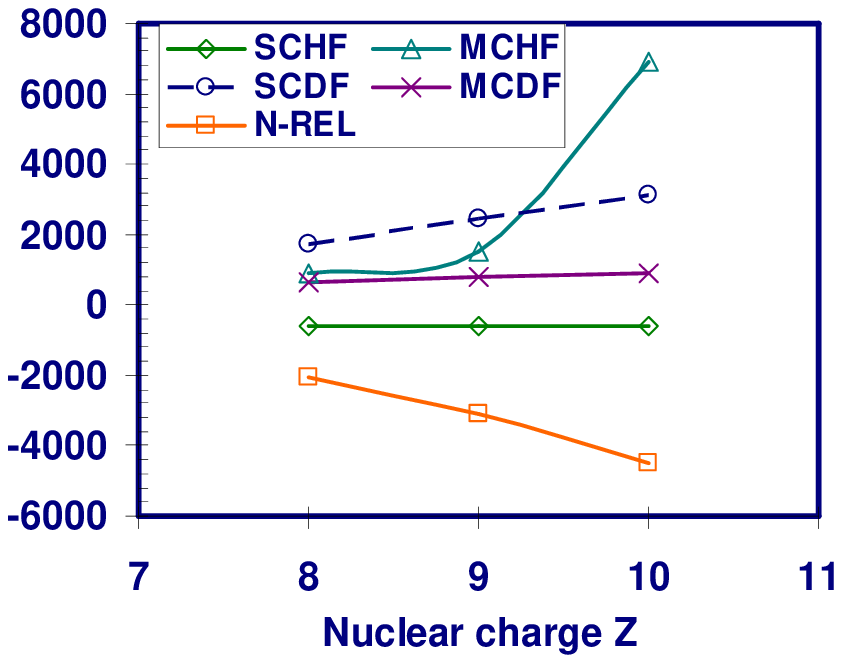}}
\caption{Difference (in cm$^{-1}$) between theoretical and
experimental transition energies for the 1s2s2p$^{2}$3d $^{6}$P -
1s2p$^{3}$3d $^{6}$D$^{o}$ transitions.} \label{fig6}
\end{figure}

We have studied differences between experimental and theoretical
transition energies of the 1s2s2p$^{2}$3d $^{6}$L - 1s2p$^{3} $3d
$^{6}$D$^{o}$ transitions along B I isoelectronic sequence. In
Fig. 4, 5 and 6 are plots of differences between theoretical and
experimental transition energies of the 1s2s2p$^{2}$3d $^{6}$L - 1s2p$^{3}$%
3d $^{6}$D$^{o}$, L=F, D, P transitions in boron-like ions. Here
theoretical transition energy is the center of gravity of the
1s2s2p$^{2}$3d $^{6}$L$_{J}$-1s2p$^{3}$3d $^{6}$D$_{J\prime }^{o}$
transition energies (computed from calculated fine structure lines
by this work) with results of theoretical analysis, and
experimental transition energy is the center
of gravity of the 1s2s2p$^{2}$3d $^{6}$L$_{J}$-1s2p$^{3}$3d $^{6}$D$%
_{J\prime }^{o}$ transition energies (computed from observed
lines) with results of experimental analysis. In Fig. 4 MCDF, SCHF
and SCDF
differences are constant for the 1s2s2p$^{2}$3d $^{6}$F - 1s2p$^{3}$3d $^{6}$D%
$^{o}$ transitions with nuclear charge Z = 8, 9 and 10.
Non-relativistic differences (N-REL) are linear. MCHF differences
for oxygen and fluorine are small, just 106 and 249 cm$^{-1}$.

In Fig. 5 SCHF, SCDF and MCDF differences are constant for the
1s2s2p$^{2} $3d $^{6}$D - 1s2p$^{3}$3d $^{6}$D$^{o}$ transitions
with nuclear charge Z = 8, 9 and 10. Non-relativistic differences
are linear. MCHF differences for oxygen and fluorine are small,
just 235 and 104 cm$^{-1}$.

In Fig. 6 SCHF and MCDF differences are constant for the 1s2s2p$^{2}$3d $%
^{6}$P - 1s2p$^{3}$3d $^{6}$D$^{o}$ transitions with nuclear
charge Z = 8, 9 and 10. SCDF and non-relativistic differences are
linear. These linear or constant energy differences can be used to
predict easily and with high accuracy transition energies for the
1s2s2p$^{2}$3d $^{6}$L - 1s2p$^{3}$3d $^{6}$D$^{o}$, L=F, D, P
transitions
for boron-like ions with 5%
\mbox{$<$}%
Z%
\mbox{$<$}%
13.
\begin{figure}[tbp]
\centerline{\includegraphics*[scale=1.00]{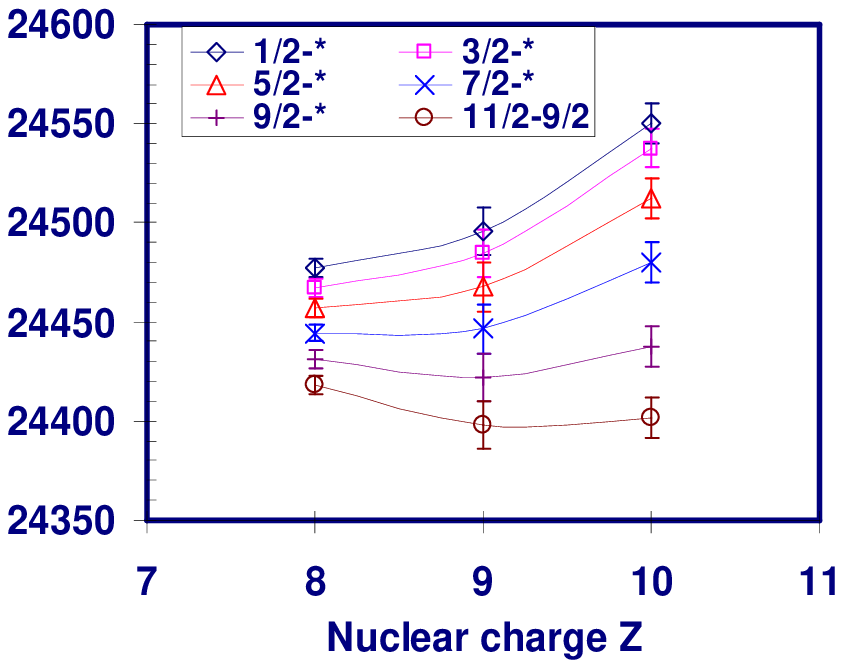}}
\caption{Reduced experimental transition energies E/(Z-1.86) (in
cm$^{-1}$) for the 1s2s2p$^{2}$3d $^{6}$F$_{J}$-1s2p$^{3}$3d
$^{6}$D$_{J\prime }^{o}$ transitions.} \label{fig7}
\end{figure}
\begin{figure}[tbp]
\centerline{\includegraphics*[scale=1.00]{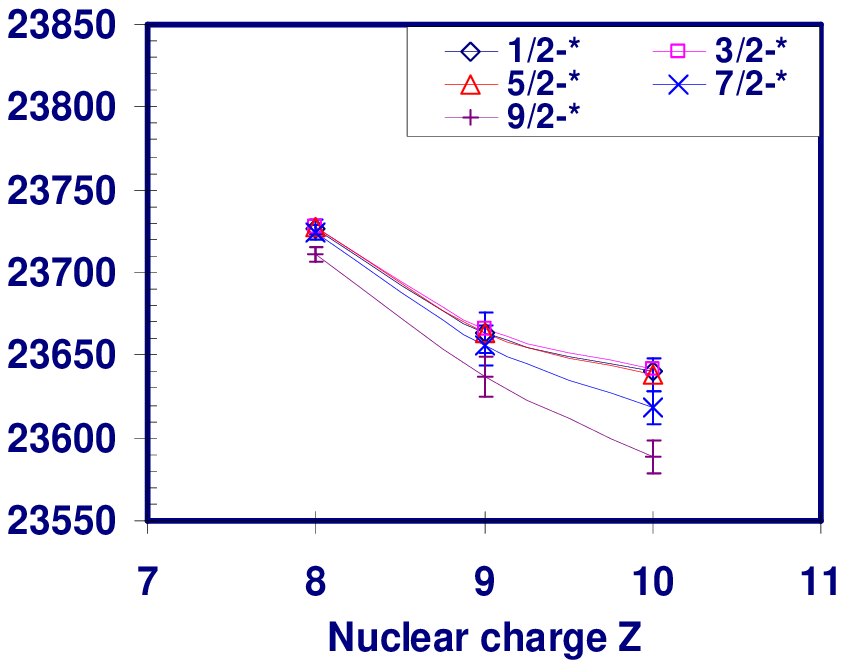}}
\caption{Reduced experimental transition energies E/(Z-1.86) (in
cm$^{-1}$) for the 1s2s2p$^{2}$3d $^{6}$D$_{J}$-1s2p$^{3}$3d
$^{6}$D$_{J\prime }^{o}$ transitions.} \label{fig8}
\end{figure}
\begin{figure}[tbp]
\centerline{\includegraphics*[scale=1.00]{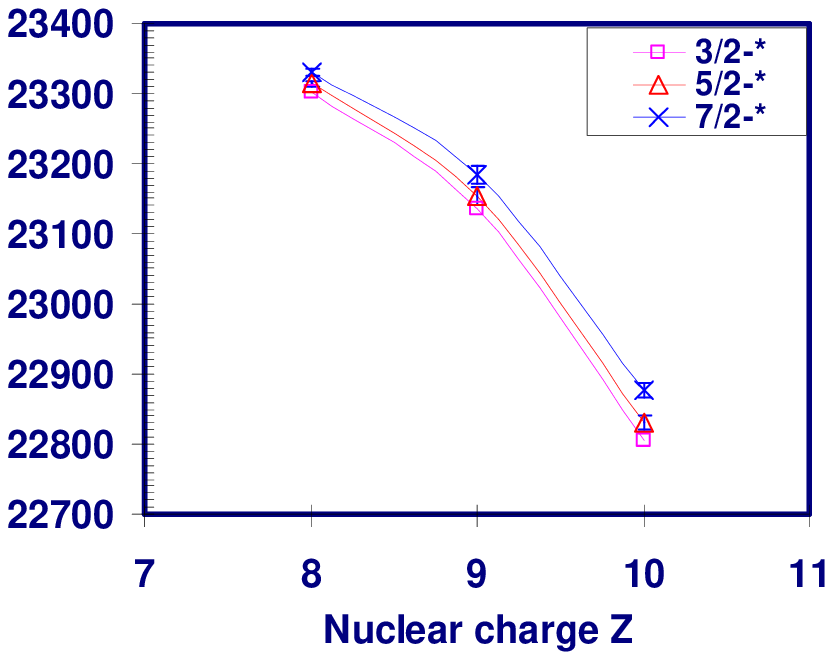}}
\caption{Reduced experimental transition energies E/(Z-1.86) (in
cm$^{-1}$) for the 1s2s2p$^{2}$3d $^{6}$P$_{J}$-1s2p$^{3}$3d
$^{6}$D$_{J\prime }^{o}$ transitions.} \label{fig9}
\end{figure}

In Fig. 7, 8 and 9 are summarized details of experimental fine
structures of the 1s2s2p$^{2}$3d $^{6}$L$_{J}$, L=F, D, P states
in O IV, F V
and Ne VI. Comparisons of measured fine structures of the 1s2s2p$^{2}$3d $%
^{6}$L$_{J}$ states show reasonable tendency. Here error bars are
from experiments.
\begin{table}
\caption{\label{tab:table1}Lifetimes (in ps) of the 1s2s2p$^{2}$3d $^{6}$P and 1s2p$^{3}$3d $%
^{6}$D$^{o}$ states in N III, O IV, F V, Ne VI and Na VII.}
\begin{ruledtabular}
\begin{tabular}{lrrrcc}
Ion & \multicolumn{3}{c}{This work} & \multicolumn{2}{c}{others} \\
\multicolumn{1}{r}{} & MCHF & SCHF & MCDF & Expt & Theory \\
\hline
\multicolumn{3}{l}{1s2s2p$^{2}$3d $^{6}$P} &  &  &  \\
N\ III & 45.37 & 40.65 & 47.42 & 66$\pm $12$^{a}$ & 42.8$\pm $18$^{b}$ \\
O\ IV & 15.32 & 15.29 & 16.01 & 12$\pm $3$^{a}$ & 14.95$\pm $4$^{b}$ \\
F V & 6.82 & 6.90 & 7.02 & 11$\pm $4$^{a}$ & 6.689$\pm $4$^{b}$ \\
Ne VI & 3.57 & 3.54 & 3.61 & \multicolumn{1}{r}{} & \multicolumn{1}{r}{} \\
Na VII & 2.10 & 2.04 & 2.07 & \multicolumn{1}{r}{} & \multicolumn{1}{r}{} \\
\multicolumn{3}{l}{1s2p$^{3}$3d $^{6}$D$^{o}$} &  &
\multicolumn{1}{r}{} &
\multicolumn{1}{r}{} \\
N\ III & 275.3 & 286.2 & 255.3 & \multicolumn{1}{r}{} &
\multicolumn{1}{r}{}
\\
O\ IV & 225.7 & 238.7 & 213.1 & \multicolumn{1}{r}{} &
\multicolumn{1}{r}{}
\\
F\ V & 196.7 & 204.3 & 183.3 & \multicolumn{1}{r}{} &
\multicolumn{1}{r}{}
\\
Ne VI & 170.1 & 178.1 & 153.0 & \multicolumn{1}{r}{} &
\multicolumn{1}{r}{}
\\
Na VII & 150.8 & 157.4 & 140.0 & \multicolumn{1}{r}{} &
\multicolumn{1}{r}{}
\end{tabular}
\end{ruledtabular}
\begin{tabular}{l}
a Blanke ~\cite{bl}. b Miecznik ~\cite{mie}.
\end{tabular}
\end{table}
\begin{figure}[tbp]
\centerline{\includegraphics*[scale=1.10]{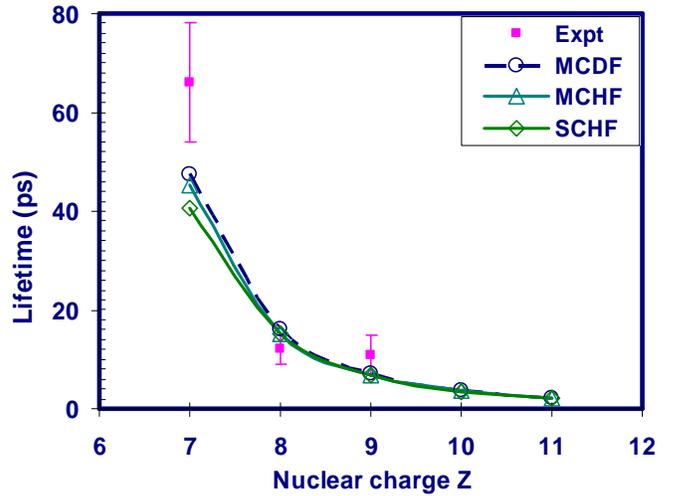}}
\caption{Lifetimes (in ps) for the 1s2s2p$^{2}$3d $^{6}$P$^{e}$
states in B I isoelectronic sequence. Experimental values are
taken from Table V. } \label{fig10}
\end{figure}
\begin{figure}[tbp]
\centerline{\includegraphics*[scale=1.10]{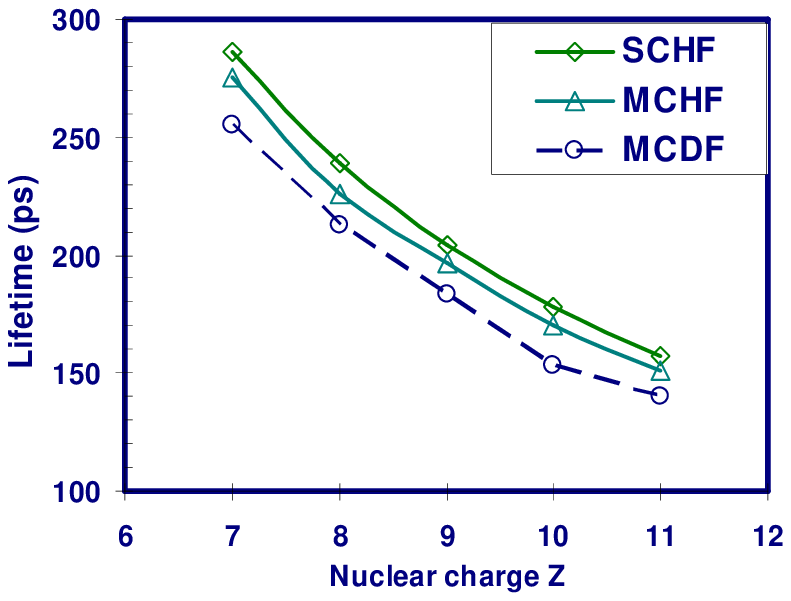}}
\caption{Lifetimes (in ps) for the 1s2p$^{3}$3d $^{6}$D$^{o}$
states in the boron sequence.} \label{fig11}
\end{figure}

MCHF, SCHF and MCDF calculated lifetimes for the 1s2s2p$^{2}$3d $%
^{6}$P and 1s2p$^{3}$3d $^{6}$D$^{o}$ states in N III, O IV, F V
Ne VI and Na VII by this work are listed in Table V, and compared
with results of measurements of Blanke et al ~\cite{bl} and
calculations of Miecznik  et al~\cite{mie}. They are plotted in
Fig. 10 and 11. Discrepancy between theory and experiments is
around or larger than experimental errors (see Fig. 10), most
probably due to additional decay modes of M2 and radiative
autoionization or some missing configurations which are important
for MCHF and MCDF calculations.
\begin{figure}[tbp]
\centerline{\includegraphics*[scale=1.2]{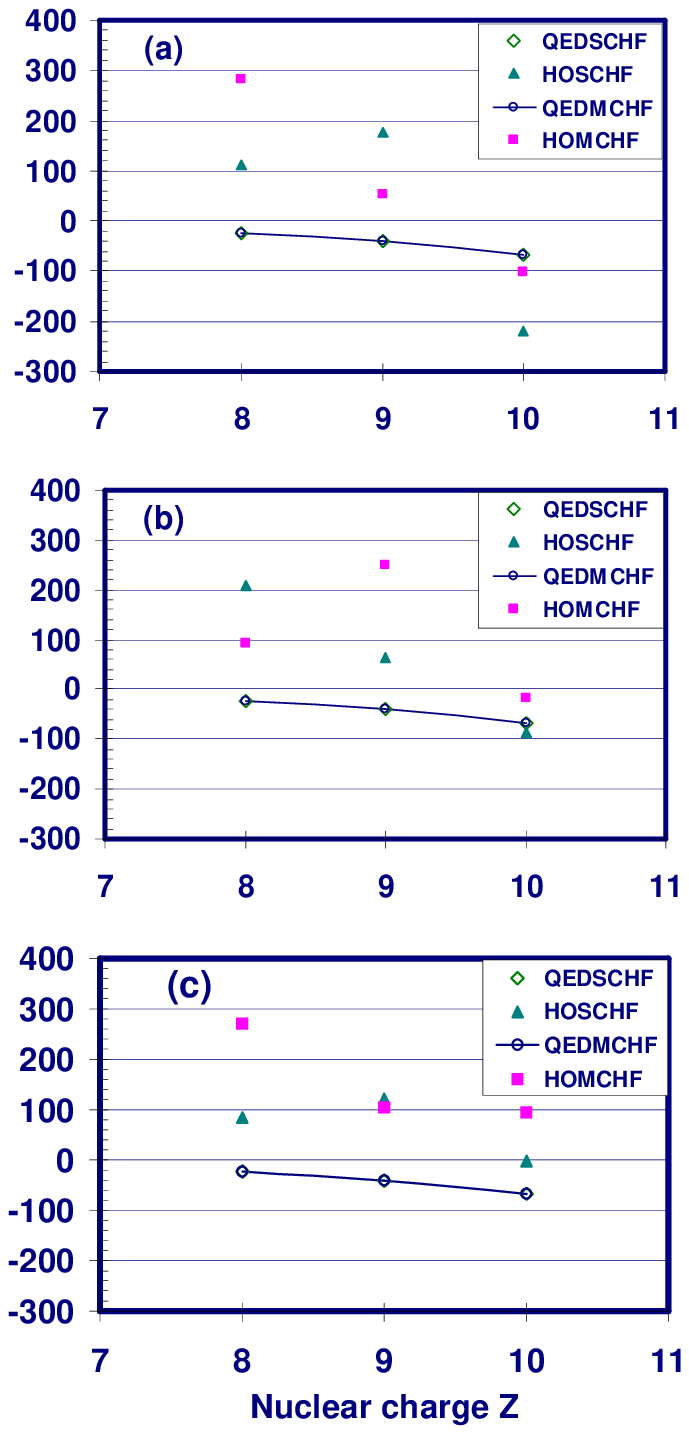}}
\caption{QED and higher-order corrections (in cm$^{-1}$) for (a) the 1s2s2p$^{2}$3d $%
^{6}$F - 1s2p$^{3}$3d $^{6}$D$^{o}$, (b) 1s2s2p$^{2}$3d $^{6}D$ - 1s2p$^{3}$%
3d $^{6}$D$^{o}$ and (c) 1s2s2p$^{2}$3d $^{6}$P - 1s2p$^{3}$3d
$^{6}$D$^{o}$ transitions in B I isoelectronic sequence.}
\label{fig12}
\end{figure}

QED and higher-order corrections for the 1s2s2p$^{2}$3d $^{6}$L - 1s2p$%
^{3}$3d $^{6}$D$^{o},$ L=F, D, P transitions in O IV, F V and Ne
VI are up to -220-370 cm-1 (see Table I, III and IV) and can't be
ignored in careful comparison with experiments. Here QED and
higher-order corrections were calculated from effective nuclear
charge Z$_{eff}$ obtained from MCHF and SCHF calculations. In Fig.
11 are plots of the above corrections to weighted mean transition
energies. The results show that
weighted mean wavelengths for the 1s2s2p$^{2}$3d $^{6}$L - 1s2p$^{3}$3d $%
^{6}$D$^{o},$ L=F, D, P transitions in O IV, F V and Ne VI are
sensitive to QED and higher-order corrections to 0.27 \AA , 0.26
\AA\ and 0.18 \AA ,
respectively. They are larger than estimated experimental precision of $%
\pm $0.06 \AA , $\pm $0.10 \AA\ and $\pm $0.05 \AA . Transition
energies are strongly related to electron correlation. We could
not get exact electron correlation. QED and higher-order
corrections of sextet states in boron-like systems are large
enough to be seen experimentally. This work could provide a good
test for QED and higher-order corrections if electron correlation
were known.

\section{CONCLUSIONS}
We performed MCHF (with QED and higher-order corrections) and MCDF
calculations to get energies, lifetimes and relevant E1 transition
wavelengths and rates for the 1s2s2p$^{2}$3d $^{6}$L -
1s2p$^{3}$3d $^{6}$D$^{o}$ , L=F, D, P electric-dipole transitions
in five-electron O IV, F V and Ne VI. Present beam-foil study of
oxygen, fluorine and neon led to observations of 42 new lines in
the sextet system of O IV, F V and Ne VI. We measured wavelengths
with good accuracy. Identifications are mainly obtained by
comparing transition wavelengths and rates with results of MCHF
and MCDF calculations. Theoretical and experimental transition
energies are consistent in errors. Differences between theoretical
and experimental transition energies are in reasonable range. For
lifetimes of the 1s2s2p$^{2}$3d $^{6}$P states in O IV and F V
there remain large discrepancies of about 20\% between results of
MCHF and MCDF calculations and experiments from~\cite{bl}.



\begin{thebibliography}{11}
\expandafter\ifx\csname natexlab\endcsname\relax\def\natexlab#1{#1}\fi
\expandafter\ifx\csname bibnamefont\endcsname\relax
  \def\bibnamefont#1{#1}\fi
\expandafter\ifx\csname bibfnamefont\endcsname\relax
  \def\bibfnamefont#1{#1}\fi
\expandafter\ifx\csname citenamefont\endcsname\relax
  \def\citenamefont#1{#1}\fi
\expandafter\ifx\csname url\endcsname\relax
  \def\url#1{\texttt{#1}}\fi
\expandafter\ifx\csname urlprefix\endcsname\relax\def\urlprefix{URL }\fi
\providecommand{\bibinfo}[2]{#2}
\providecommand{\eprint}[2][]{\url{#2}}

\bibitem[{\citenamefont{Blanke et~al}(1992)\citenamefont{Blanke, Fricke, Heckmann and Tr\"{a}bert}}]{bl}
\bibinfo{author}{\bibfnamefont{J.~H.} \bibnamefont{Blanke}},
  \bibinfo{author}{\bibfnamefont{B.} \bibnamefont{Fricke}},
  \bibinfo{author}{\bibfnamefont{P.~H.} \bibnamefont{Heckmann}}
  \bibnamefont{and} \bibinfo{author}{\bibfnamefont{E.} \bibnamefont{Tr\"{a}bert}},
  \bibinfo{journal}{Phys. Scr.
{\bf{45}}, 430}  (\bibinfo{year}{1992}).

\bibitem[{\citenamefont{Lapierre~and~Knystautas}(2000)\citenamefont{Lapierre, Knystautas}}]{lap}
\bibinfo{author}{\bibfnamefont{L.} \bibnamefont{Lapierre}}
  \bibnamefont{and} \bibinfo{author}{\bibfnamefont{E.~J.} \bibnamefont{Knystautas}},
  \bibinfo{journal}{J. Phys. B {\bf{33}}, 2245}  (\bibinfo{year}{2000}).

\bibitem[{\citenamefont{Lin and berry ~et al}(2003)\citenamefont{Lin, Berry, Shibata, Livingaton,
Garnir, Bastin, D\'{e}sesquelles, Savukov}}]{lin}
\bibinfo{author}{\bibfnamefont{Bin} \bibnamefont{Lin}},
\bibinfo{author}{\bibfnamefont{H. Gordon} \bibnamefont{Berry}},
  \bibnamefont{and} \bibinfo{author}{\bibfnamefont{Tomohiro} \bibnamefont{Shibata}},
\bibinfo{author}{\bibfnamefont{A. E.} \bibnamefont{Livingston}},
\bibinfo{author}{\bibfnamefont{Henri-Pierre} \bibnamefont{Garnir}},
\bibinfo{author}{\bibfnamefont{Thierry} \bibnamefont{Bastin}},
\bibinfo{author}{\bibfnamefont{J.} \bibnamefont{D\'{e}sesquelles}},
\bibinfo{author}{\bibfnamefont{Igor} \bibnamefont{Savukov}},
  \bibinfo{journal}{Phys. Rev. A {\bf{67}}, 062507}  (\bibinfo{year}{2003}).

\bibitem[{\citenamefont{Berry ~et al}(1975)\citenamefont{Berry}}]{berry1}
\bibinfo{author}{\bibfnamefont{H. G. } \bibnamefont{Berry}},
\bibinfo{author}{\bibfnamefont{T.} \bibnamefont{Bastin}},
\bibinfo{author}{\bibfnamefont{E.} \bibnamefont{Biemont}},
\bibinfo{author}{\bibfnamefont{P.~D.} \bibnamefont{ Dumont}}
  \bibnamefont{and} \bibinfo{author}{\bibfnamefont{H.~P.} \bibnamefont{Garnir}},
  \bibinfo{journal}{Rep. Prog. Phys. {\bf{5}}, 12}  (\bibinfo{year}{1975}).

\bibitem[{\citenamefont{Kramida ~et al}(1999)\citenamefont{Kramida, Bastin, Biemont, Dumont, Garnir}}]{kla}
\bibinfo{author}{\bibfnamefont{A.~ E.} \bibnamefont{Kramida}},
\bibinfo{author}{\bibfnamefont{T.} \bibnamefont{Bastin}},
\bibinfo{author}{\bibfnamefont{E.} \bibnamefont{Biemont}},
\bibinfo{author}{\bibfnamefont{P.~D.} \bibnamefont{ Dumont}}
  \bibnamefont{and} \bibinfo{author}{\bibfnamefont{H.~P.} \bibnamefont{Garnir}},
  \bibinfo{journal}{J. Opt. Soc. Am. B {\bf{16}} (11), 1966}  (\bibinfo{year}{1999}).

\bibitem[{\citenamefont{Miecznik ~et al}(2000)\citenamefont{Miecznik, Brage,Fischer}}]{mie}
\bibinfo{author}{\bibfnamefont{G.} \bibnamefont{Miecznik}},
\bibinfo{author}{\bibfnamefont{T.} \bibnamefont{Brage}}
  \bibnamefont{and} \bibinfo{author}{\bibfnamefont{C.~F.} \bibnamefont{Fischer}},
  \bibinfo{journal}{Phys. Scr. {\bf{45}}, 436}  (\bibinfo{year}{1992}).

\bibitem[{\citenamefont{Fischer ~et al}(2000)\citenamefont{Fischer, Brage, Jonsson}}]{fibk}
\bibinfo{author}{\bibfnamefont{C.~F.} \bibnamefont{Fischer}},
\bibinfo{author}{\bibfnamefont{T.} \bibnamefont{Brage}}
  \bibnamefont{and} \bibinfo{author}{\bibfnamefont{P.} \bibnamefont{Jonsson}},
  \bibinfo{journal}{Computational Atomic Structure an
MCHF Approach (Institute of Physics Publishing, Bristal and Philadelphia}  (\bibinfo{year}{1997}).

\bibitem[{\citenamefont{Chung ~et al}(1992)\citenamefont{Chung, Zhu, Wang}}]{kt1}
\bibinfo{author}{\bibfnamefont{K. T.} \bibnamefont{Chung}},
\bibinfo{author}{\bibfnamefont{X. W.} \bibnamefont{Zhu}}
  \bibnamefont{and} \bibinfo{author}{\bibfnamefont{Z. W. } \bibnamefont{Wang}},
  \bibinfo{journal}{Phys. Rev. A {\bf{29}}, 682}  (\bibinfo{year}{1984}).

\bibitem[{\citenamefont{Dyall~&~ Grant ~et al}(1989)\citenamefont{Dyall, Grant }}]{MCDF}
\bibinfo{author}{\bibfnamefont{K.~G.} \bibnamefont{Dyall}},
  \bibnamefont{and} \bibinfo{author}{\bibfnamefont{I.~P.} \bibnamefont{Grant}},
  \bibinfo{journal}{computer physics communications {\bf{55}}, 425}  (\bibinfo{year}{1989}).

\bibitem[{\citenamefont{Parpia, Fischer and Grant}(1996)\citenamefont{Parpia, Fischer, Grant}}]{MCDF1}
\bibinfo{author}{\bibfnamefont{F.~A.} \bibnamefont{Parpia}},
\bibinfo{author}{\bibfnamefont{C.~F.} \bibnamefont{Fischer}},
  \bibnamefont{and} \bibinfo{author}{\bibfnamefont{I.~P.} \bibnamefont{Grant}},
  \bibinfo{journal}{Computer Physics Communications {\bf{94}} (2-3), 249}  (\bibinfo{year}{1996}).

\bibitem[{\citenamefont{Fritzsche~and~Grant~&~ Grant ~et al}(1997)\citenamefont{Fritzsche and Grant}}]{MCDF2}
\bibinfo{author}{\bibfnamefont{S.} \bibnamefont{Fritzsche}},
  \bibnamefont{and} \bibinfo{author}{\bibfnamefont{I.~P.} \bibnamefont{Grant}},
  \bibinfo{journal}{Computer Physics Communications {\bf{103}} (2-3), 277}  (\bibinfo{year}{1997}).



\bibitem[{\citenamefont{Chung ~et al}(1992)\citenamefont{Chung, Zhu, Wang}}]{qed1}
\bibinfo{author}{\bibfnamefont{K. T.} \bibnamefont{Chung}},
\bibinfo{author}{\bibfnamefont{X. W.} \bibnamefont{Zhu}}
  \bibnamefont{and} \bibinfo{author}{\bibfnamefont{Z. W. } \bibnamefont{Wang}},
  \bibinfo{journal}{Phys. Rev. A {\bf{47}} (3) 1740}  (\bibinfo{year}{1992}).

\bibitem[{\citenamefont{Chung ~et al}(1993)\citenamefont{Chung, Zhu}}]{qed2}
\bibinfo{author}{\bibfnamefont{K. T. } \bibnamefont{Chung}}
  \bibnamefont{and} \bibinfo{author}{\bibfnamefont{X. W. } \bibnamefont{Zhu}},
  \bibinfo{journal}{Phys. Rev. A {\bf{48}}(3) 1944}  (\bibinfo{year}{1993}).

\bibitem[{\citenamefont{Drake~and~Swainson}(1990)\citenamefont{Drake, Swainson}}]{drake}
\bibinfo{author}{\bibfnamefont{G. W. F. } \bibnamefont{Drake}}
  \bibnamefont{and} \bibinfo{author}{\bibfnamefont{R. A. } \bibnamefont{Swainson}},
  \bibinfo{journal}{Phys. Rev. A {\bf{41}} (3) 1243}  (\bibinfo{year}{1990}),


\bibitem[{\citenamefont{Berry ~et al}(1982)\citenamefont{Berry, Brooks, Cheng, Hardis, Ray}}]{berry2}
\bibinfo{author}{\bibfnamefont{H.~G.} \bibnamefont{Berry}},
\bibinfo{author}{\bibfnamefont{R.~L.} \bibnamefont{Brooks}},
\bibinfo{author}{\bibfnamefont{K.~T} \bibnamefont{Cheng}},
\bibinfo{author}{\bibfnamefont{J.~E.} \bibnamefont{Hardis}}
  \bibnamefont{and} \bibinfo{author}{\bibfnamefont{W.} \bibnamefont{Ray}},
  \bibinfo{journal}{Phys.
Scr. {\bf{42}}, 391}  (\bibinfo{year}{1982}).

\bibitem[{\citenamefont{Hardis ~et al}(1984)\citenamefont{Hardis, Berry, Curtis and Livingston}}]{neq}
\bibinfo{author}{\bibfnamefont{J.~E.} \bibnamefont{Hardis}},
\bibinfo{author}{\bibfnamefont{H.~G.} \bibnamefont{Berry}},
\bibinfo{author}{\bibfnamefont{L.~G.} \bibnamefont{Curtis}}
  \bibnamefont{and} \bibinfo{author}{\bibfnamefont{A.~E} \bibnamefont{Livingston}},
  \bibinfo{journal}{Phys.
Scr. {\bf{30}}, 189}  (\bibinfo{year}{1984}).

\bibitem[{\citenamefont{Garnir ~et al}(1988)\citenamefont{Garnir, Baudinet-Robinet and Dumont}}]{gar}
\bibinfo{author}{\bibfnamefont{H.~P.} \bibnamefont{Garnir}},
\bibinfo{author}{\bibfnamefont{Y.} \bibnamefont{Baudinet-Robinet}},
  \bibnamefont{and} \bibinfo{author}{\bibfnamefont{P.~D.} \bibnamefont{ Dumont}},
  \bibinfo{journal}{Nuclear Instruments
and Methods in Physics Research B {\bf{31}}, 161}  (\bibinfo{year}{1988}).


\bibitem[{\citenamefont{Baudinet-Robinet ~et al}(1990)\citenamefont{Baudinet-Robinet, Dumont and Garnir, }}]{rob}
\bibinfo{author}{\bibfnamefont{Y.} \bibnamefont{Baudinet-Robinet}},
  \bibnamefont{and} \bibinfo{author}{\bibfnamefont{P.~D.} \bibnamefont{ Dumont}},
  \bibinfo{author}{\bibfnamefont{H.~P.} \bibnamefont{Garnir}},
  \bibinfo{journal}{PhysicaliaMag. {\bf{12}}, 3}  (\bibinfo{year}{1990}).

\bibitem[{\citenamefont{Girardeau ~et al}(1971)\citenamefont{Girardeau, Knystautas, Beauchemin sj, Neveu, Drouin}}]{gir}
\bibinfo{author}{\bibfnamefont{R.} \bibnamefont{Girardeau}}
  \bibnamefont{and} \bibinfo{author}{\bibfnamefont{E.~J.} \bibnamefont{Knystautas}},
  \bibinfo{author}{\bibfnamefont{G.} \bibnamefont{Beauchemin sj}}
  \bibinfo{author}{\bibfnamefont{B.} \bibnamefont{Neveu}}
  \bibinfo{author}{\bibfnamefont{R.} \bibnamefont{Drouin}}
  \bibinfo{journal}{J. Phys. B {\bf{4}}, 1743}  (\bibinfo{year}{1971}).

\bibitem[{\citenamefont{Lin and berry ~et al}(2003)\citenamefont{Lin, Berry, Shibata, Livingaton,
Garnir, Bastin, D\'{e}sesquelles}}]{lin1}
\bibinfo{author}{\bibfnamefont{Bin} \bibnamefont{Lin}},
\bibinfo{author}{\bibfnamefont{H. Gordon} \bibnamefont{Berry}},
  \bibnamefont{and} \bibinfo{author}{\bibfnamefont{Tomohiro} \bibnamefont{Shibata}},
\bibinfo{author}{\bibfnamefont{A. E.} \bibnamefont{Livingston}},
\bibinfo{author}{\bibfnamefont{Henri-Pierre} \bibnamefont{Garnir}},
\bibinfo{author}{\bibfnamefont{Thierry} \bibnamefont{Bastin}},
\bibinfo{author}{\bibfnamefont{J.} \bibnamefont{D\'{e}sesquelles}},
  \bibinfo{journal}{sumitted to Phys. Rev. A {\bf{}}, }  (\bibinfo{year}{2004}).
  \eprint{arXiv:physics/0404001}.

\bibitem[{\citenamefont{Bockasten~and~Johansson}(1968)\citenamefont{Bockasten, Johansson}}]{o1}
\bibinfo{author}{\bibfnamefont{K.} \bibnamefont{Bockasten}}
  \bibnamefont{and} \bibinfo{author}{\bibfnamefont{K.~B.} \bibnamefont{Johansson}},
  \bibinfo{journal}{Ark. Fys. {\bf{38}}, 563}  (\bibinfo{year}{1968}).

\bibitem[{\citenamefont{Pettersson}(1982)\citenamefont{Pettersson}}]{o2}
\bibinfo{author}{\bibfnamefont{S.~G.} \bibnamefont{Pettersson}},
  \bibinfo{journal}{Phys. Scr. {\bf{26}}, 296}  (\bibinfo{year}{1982}).

\bibitem[{\citenamefont{Moore}(1965-1983)\citenamefont{Moore}}]{o3}
\bibinfo{author}{\bibfnamefont{C.~E} \bibnamefont{Moore}},
  \bibinfo{journal}{NSRDS-NBS {\bf{3}}, Section 1-10}  (\bibinfo{year}{1965-1983}).

\bibitem[{\citenamefont{Moore}(1949)\citenamefont{Moore}}]{f1}
\bibinfo{author}{\bibfnamefont{C.~E} \bibnamefont{Moore}},
  \bibinfo{journal}{Atomic Energy Levels {\bf{1}}, Circ. Natl. Bur. Stand. 467}  (\bibinfo{year}{1949}).

\bibitem[{\citenamefont{Engstr\"{o}m}(1985)\citenamefont{Engstr\"{o}m}}]{f2}
\bibinfo{author}{\bibfnamefont{L.} \bibnamefont{Engstr\"{o}m}},
  \bibinfo{journal}{Phys. Scr. {\bf{29}}, 113}  (\bibinfo{year}{1985}).

\bibitem[{\citenamefont{Brown}(1969)\citenamefont{Brown}}]{ne1}
\bibinfo{author}{\bibfnamefont{R.~T.} \bibnamefont{Brown}},
  \bibinfo{journal}{APJ {\bf{158}}, 829}  (\bibinfo{year}{1969}).

\bibitem[{\citenamefont{Vainshtein~and~Safronova}(1985)\citenamefont{Vainshtein, Safronova}}]{ne2}
\bibinfo{author}{\bibfnamefont{L.~A.} \bibnamefont{Vainshtein}}
  \bibnamefont{and} \bibinfo{author}{\bibfnamefont{U.~I.} \bibnamefont{Safronova}},
  \bibinfo{journal}{Phys. Scr. {\bf{31}}, 519}  (\bibinfo{year}{1985}).




\end{thebibliography}

\end{document}